\documentclass[aps,prb,preprint,amsmath,amssymb,endfloats]{revtex4}
\usepackage[sort&compress]{natbib}
\usepackage{graphicx}
\usepackage{gensymb}
\usepackage{xcolor}
\usepackage{float}

\begin{document}

\title{Folding of Protein L with implications for collapse in the denatured state ensemble}
\author{Hiranmay Maity}
\author{Govardhan Reddy}
\email{greddy@sscu.iisc.ernet.in}
\affiliation{Solid State and Structural Chemistry Unit, Indian Institute of Science, Bangalore, Karnataka, India 560012}

\begin{abstract} 
A fundamental question in protein folding is whether the coil to globule collapse transition occurs during the initial stages of folding (burst-phase) or simultaneously with the protein folding transition. Single molecule fluorescence resonance energy transfer (FRET) and small angle X-ray scattering (SAXS) experiments disagree on whether Protein L collapse transition occurs during the burst-phase of folding. We study Protein L folding using a coarse-grained model and molecular dynamics simulations. The collapse transition in Protein L is found to be concomitant with the folding transition. In the burst-phase of folding, we find that FRET experiments overestimate radius of gyration, $R_g$, of the protein due to the application of Gaussian polymer chain end-to-end distribution to extract $R_g$ from the FRET efficiency. FRET experiments estimate $\approx$ 6\AA \ decrease in $R_g$  when the actual decrease is $\approx$ 3\AA \ on Guanidinium Chloride denaturant dilution from 7.5M to 1M, and thereby suggesting pronounced compaction in the protein dimensions in the burst-phase. The $\approx$ 3\AA \ decrease is close to the statistical uncertainties of the $R_g$ data measured from SAXS experiments, which suggest no compaction, leading to a disagreement with the FRET experiments. The transition state ensemble (TSE) structures in Protein L folding are globular and extensive in agreement with the $\Psi$-analysis experiments. The results support the hypothesis that the TSE of single domain proteins depend on protein topology, and are not stabilised by local interactions alone.
\end{abstract} 

\maketitle

\newpage
 
\section*{\Large{Introduction}}
 
There is an ongoing debate\cite{Ziv09PCCP,Sosnick11COSB,Haran12COSB,Thirumalai13COSB,Udgaonkar13ABB} on whether the denatured ensemble of single domain proteins undergoes a coil to globule transition during the burst-phase of folding as the denaturant concentration is diluted to lower values. Proteins are heteropolymers and behave like random coils at high temperatures or denaturant concentrations\cite{Tanford66JBC,Kohn04PNAS,Hofmann12PNAS}. An interesting question is whether proteins akin to polymers undergo a collapse transition in the burst-phase of folding as the conditions are made conducive for folding\cite{Bryngelson90Biopolymer,Chan91ARBBC,OBrien08PNAS}. Single domain proteins unlike polymers are finite sized, and are composed of a specific sequence of amino acids which are hydrophobic and hydrophilic in character. The finite size effects and heteropolymer character are the reasons attributed to the marginal stability of proteins, and the near overlap of the collapse and folding transition temperatures, which makes them fold efficiently\cite{Camacho93PNAS,Li04PRL}.

The collapse transition in proteins is generally studied using single molecule fluorescence resonance energy transfer (FRET), and small angle X-ray scattering (SAXS) experiments. Although FRET and SAXS experiments agree that proteins like Cytochrome {\it c}\cite{Kathuria14JMB,Lapidus07BiophysicsJ,Hagen00JMB,Shastry98ACS} and Monellin\cite{Reddy15Biochemistry,Jha09PNAS,Kimura05PNAS} collapse during the burst-phase of folding, their results disagree for Protein L. FRET experiments for Protein L\cite{Sherman06PNAS, Merchant07PNAS,Waldauer08HFSP} infer collapse, whereas SAXS experiments\cite{Plaxco99NSB,Yoo12JMB} conclude no collapse in the burst-phase of folding on dilution of Guanidine Hydrochloride $[GuHCl]$. Both FRET and SAXS estimate the radius of gyration, $R_g$, of the protein to infer the size of the protein in the unfolded ensemble. The difference in the $R_g$ predictions of  FRET\cite{Sherman06PNAS, Merchant07PNAS} and SAXS\cite{Yoo12JMB,Plaxco99NSB} experiments for Protein L during the burst-phase of folding is statistically significant. The reasons for the disagreement between these experiments for Protein L are not completely clear. Understanding the impact of various approximations used in these methods to estimate the size of the protein can not only aid in resolving the disagreement between FRET and SAXS but also to understand the problem of protein collapse better.

Single domain proteins close to the melting temperature or the mid-point denaturant concentration generally fold in a 2-state manner through an ensemble of transition state structures (TSE). $\phi$-analysis experiments for Protein L\cite{Scalley97Biochemistry,Gu97JMB,Kim98JMB,Kim00JMB} predict that the TSE is polarised with only the N-terminal $\beta$-hairpin present. Whereas the $\psi$-analysis experiments\cite{Yoo12JMB2} predict that the TSE is globular and extensive with both the N and C-termini $\beta$-hairpins present along with some non-native interactions in the C-terminal $\beta$-hairpin. The $\psi$-analysis experiments support that the TSE and the folding pathways for the single domain proteins depend on the protein topology\cite{Baxa08JMB}, whereas the $\phi$-analysis experiments conclude that the TSE is mostly stabilised by local interactions\cite{Naganathan10PNAS}.  

Various aspects of Protein L folding such as folding pathways, transition state structures, and properties of the unfolded ensemble are studied\cite{Koga01JMB, Karanicolas02ProtSci, Clementi03JMB, Ejtehadi04PNAS, Brown04ProtSci, Yang08BMC,Voelz10JACS,Chen14PCCP} using both coarse-grained and atomistic simulations. In this manuscript, we study the burst-phase folding of Protein L to understand the origin of discrepancy between the FRET\cite{Sherman06PNAS, Merchant07PNAS} and SAXS\cite{Plaxco99NSB, Yoo12JMB} experiments using the native-centric self-organised polymer model with side chains (SOP-SC)\cite{Hyeon06Structure, Liu11PNAS} and molecular dynamics simulations. The effect of $[GuHCl]$ on Protein L conformations is taken into account using the molecular transfer model (MTM)\cite{OBrien08PNAS,Liu12JPCB}.

The computed FRET efficiency, $\langle E \rangle$, for Protein L in the burst-phase of folding is in quantitative agreement with the FRET experiments of Eaton {\it et al.}\cite{Merchant07PNAS}, and only in partial agreement with the experiments of Haran {\it et al.}\cite{Sherman06PNAS}. The FRET experiments are found to overestimate $R_g$ compared to the actual values computed directly from the simulations owing to the use of the Gaussian polymer chain end-to-end distribution function to extract $R_g$ from $\langle E \rangle$. The deviation between FRET-extracted and actual $R_g$ increased with  $[GuHCl]$. As a result, FRET experiments\cite{Merchant07PNAS} estimate $\approx6$\AA \  decrease in $R_g$ and infer protein collapse in the burst-phase, when the actual decrease is $\approx3$\AA \ as $[GuHCl]$ is diluted from 7.5M to 1M. 

The equilibrium $R_g$ computed as a function of $[GuHCl]$ is in near quantitative agreement with the SAXS experiments\cite{Yoo12JMB}. The SAXS experiments\cite{Yoo12JMB} infer no protein collapse as the burst-phase $R_g$ at $[GuHCl]$ = 4.0M and 0.67M are not statistically different. In the simulations, the burst-phase $R_g$ at $[GuHCl]$ = 4M and 1M are $25.1\pm4.0$\AA \ and $23.8\pm3.9$\AA, respectively, a difference of $\approx 1.3$\AA, which is well within the standard deviation $\sigma_{R_g}\approx$ 4\AA. From this analysis, which is similar to the SAXS analysis, we can infer no collapse as the change in protein dimensions are not statistically significant, leading to a disagreement with the FRET experiments.

The TSE of Protein L at the melting temperature is inferred using $P_{fold}$ calculations\cite{Du98JCP}. The TSE is found to be globular and extensive with both the N and C termini $\beta$-hairpins present resembling a topology similar to that of the folded structure. The results are in agreement with the $\Psi$-analysis experiments\cite{Yoo12JMB2} and only in partial agreement with the $\phi$-analysis experiments\cite{Scalley97Biochemistry,Gu97JMB,Kim98JMB,Kim00JMB}. The inferred TSE support the hypothesis that the transition state structures of single domain homologous proteins are extensive and depend on the protein topology\cite{Baxa15PNAS}. 

 \section*{\Large{Methods}}

{\bf Self Organised polymer-Side Chain (SOP-SC) Model:} We used the SOP-SC (self-organized polymer-side chain) model\cite{Hyeon06Structure,Liu11PNAS} in which each amino acid residue is represented by two beads. One bead is at the C$_\alpha$ position representing the backbone atoms, and the other bead is at the center of mass of the side chain representing the side chain atoms. The effective energy of a protein conformation in the SOP-SC model is a sum of bonded and non-bonded interactions. The bonded interactions ($E_B$) are present between a pair of  connected beads. The non-bonded interactions are a sum of native ($E_{NB}^{N}$) and non-native ($E_{NB}^{NN}$) interactions (see Supporting Information (SI) for more details).  The native interactions for protein L are identified using the crystal structure\cite{Neill01ActaCry} (Protein Data Bank ID: 1HZ6) (Fig 1A). The number of residues in the crystal structure, $N_{res}=64$. The native interactions between the beads representing the amino acid side chains interact via a residue dependent Betancourt-Thirumalai statistical potential\cite{Betancourt99ProtSci}. 

The coarse-grained force-field in the SOP-SC model for a protein conformation given by the co-ordinates $\{\bf r\}$ in the absence of denaturants, $[C]=0$, is 
\begin{equation}\label{pot}
E_{CG}(\{{\bf r}\},0)  = E_{B} + E_{NB}^{N} +  E_{NB}^{NN}. 
\end{equation}
Description of the various energy terms in Equation~\ref{pot} and the parameters used in the energy function are given in the SI. These parameters are identical to the values previously used to successfully study the folding properties of the proteins Ubiquitin\cite{Reddy15JPCB} and GFP\cite{Reddy12PNAS}. We used the same force-field to study the properties of different proteins, and as a result this force-field satisfies the criterion of a transferable force-field. \\
  
{\bf Molecular Transfer Model:}  To simulate Protein L folding thermodynamics and kinetics in the presence of $[GuHCl]$ we used the Molecular Transfer Model (MTM)\cite{OBrien08PNAS, Liu12JPCB}. In the presence of a denaturant of concentration $[C]$, the effective coarse-grained force field for the protein using MTM is given by
\begin{equation}\label{mtmpot}
E_{CG}(\{ {\bf r} \},[C])  = E_{CG}(\{{\bf r} \},0) + \Delta G_{tr}(\{ {\bf r} \},[C]),             
\end{equation} 
   where $E_{CG}(\{{\bf r} \},0)$ is given by Eq.~\ref{pot}, $\Delta G_{tr}(\{ {\bf r} \},[C])$ is the protein-denaturant interaction energy in a solution with denaturant concentration [C], and is given by
\begin{equation}
\Delta G_{tr}(\{ {\bf r} \},[C])  =  \sum_{k=1}^{N} \delta g_{tr,k}([C]) \alpha_k(\{ {\bf r} \})/\alpha_{Gly-k-Gly},
\label{trans_ene}
\end{equation}
where $N$(=$N_{res} \times 2 =128$) is the number of beads in coarse-grained Protein L, $\delta g_{tr,k}([C])$ is the  transfer free energy of bead $k$, $\alpha_k(\{ {\bf r} \})$ is the solvent accessible surface area (SASA) of the bead $k$ in a protein conformation described by positions  $\{\bf r\}$, $\alpha_{Gly-k-Gly}$ is the SASA of the bead $k$ in the tripeptide $Gly-k-Gly$. The radii for amino acid side chains to compute $\alpha_k(\{ {\bf r} \})$ are given in Table~S2 in Ref.\cite{Reddy15JPCB}. The experimental\cite{Auton04Biochem,OBrien08PNAS,OBrien09Biochem} transfer free energies $\delta g_{tr,i}([C])$, which depend on the chemical nature of the denaturant, for backbone and side chains are listed in Table S3 in Ref.\cite{Liu11PNAS}.  The values for $\alpha_{Gly-k-Gly}$ are listed in Table S4 in Ref.\cite{Liu11PNAS}. \\

{\bf Simulations and Data Analysis:}
Low friction Langevin dynamics simulations\cite{Veitshans97FoldDes} are used to generate protein conformations as a function of $T$ in $[C]=0M$ conditions. To compute thermodynamic properties of the protein in a denaturant solution of concentration $[C]$, $\Delta G_{tr}(\{ {\bf r} \},[C])$  is treated as perturbation to $E_{CG}(\{{\bf r} \},0)$ in Eq.~\ref{mtmpot}, and  Weighted Histogram Method\cite{Kumar92JCompChem,OBrien08PNAS, Liu12JPCB} is used to compute average value of various physical quantities at any [C]. Brownian dynamics simulations\cite{Ermak78JChemPhys} are used with the full Hamiltonian (Eq.~\ref{mtmpot}) to simulate the burst-phase folding kinetics of the protein in a denaturant solution of concentration $[C]$ (see SI for details).

We computed  structural overlap function\cite{Guo96JMB}, $\chi$, and radius of gyration, $R_g$, to monitor protein L folding kinetics. The structural overlap function is defined as $\chi = 1 - \frac{1}{N_{tot}} \sum\limits_{i=1}^{N_{tot}}\Theta\left(\delta-\lvert r_{i}-r_{i}^0\rvert\right)$. Here, $N_{tot} (=777)$ is the number of pairs of beads in the SOP-SC model of Protein L assuming that the bead centers are separated by at least 2 bonds, $r_{i}$ is the distance between the $i^{th}$ pair of beads, and $r_{i}^0$ being the corresponding distance in the folded state, $\Theta$ is the Heaviside step function, and $\delta = 2$\AA. Using $\chi$ as an order parameter,  we calculated the fraction of molecules in the NBA, $f_{NBA}$ as a function of $[GuHCl]$(see SI for details and Fig.~S2). $R_g$,  is calculated using $R_g = (1/2N^2)(\sum\limits_{i, j}\vec{r}_{ij}^{ \ 2})^{1/2}$, where $\vec{r}_{ij}$ is the vector connecting the beads $i$ and $j$. The extent of long-range contacts in the TSE structures compared to the coarse-grained PDB structure is analysed using the relative contact order\cite{Plaxco98JMB, Klimov98JMB}, $RCO$, which is defined as $RCO=\frac{1}{N_{res}N_{nat}}\sum\limits_{i=1}^{N_{nat}}L_i$, where $N_{nat}$ is the number of pairs of beads with native interactions in the protein conformation (see SI), and $L_i$ is the number of residues separating the contact pair $i$.\\

{\section*{\Large {Results and Discussion}} 

{\bf Thermodynamics of Protein L folding}: The protein in the folded state has one $\alpha$-helix ($\alpha_1$) and four $\beta$-strands ($\beta_1$ - $\beta_4$) (Fig.~\ref{denat}A and S1A). Low friction Langevin dynamics simulations performed at different temperatures, $T$, ranging from $300K$ to $430K$  show that folding occurs in a two-state manner (see SI, Fig.~S1). The melting temperature, $T_M$, of protein L obtained from the heat capacity, $C_v$, plot is $374.5K$ (Fig.~S1), and the value observed in experiments\cite{Gillespie00PNAS} is $348.5K$. The difference in $T_M$ between experiments and simulations can be attributed to the simplified coarse-grained SOP-SC model. At $T_M$, the protein transitions between the native basin of attraction (NBA) and the unfolded basin of attraction (UBA) (Fig.~S2). Protein L folding thermodynamics and kinetics in the presence of the denaturant $GuHCl$ is studied using the molecular transfer model (MTM)\cite{OBrien08PNAS,Liu12JPCB}. In order to compare the denaturant-dependent folding properties of the protein computed from simulations with the experiments, a simulation temperature $T_S (=357.7K)$ at which theoretically obtained free energy difference between the NBA and UBA, $\Delta G_{NU}^{Sim}(=G_N(T_S) - G_U(T_S))$ matches with the experimentally\cite{Kim00JMB} measured value, $\Delta G_{NU}^{Exp}$ (= -4.6 kcal/mole), at $[GuHCl]=0M$ is used. This is the only adjustable parameter in the model, which is equivalent to matching the energy scales between the simulations and experiments.

The structural overlap parameter, $\chi$ (see methods), is used to distinguish between the NBA and UBA protein conformations (Fig.~S2). The protein conformations with $\chi \leq 0.47$ belong to the NBA, and conformations with $\chi > 0.47$ belong to the UBA (Fig.~S2B). The fraction of molecules in NBA, $f_{NBA}$, as a function of $[GuHCl]$ computed from simulations is in  quantitative agreement with the experiments\cite{Kim00JMB,Sherman08Biophys,Waldauer08HFSP} (Fig.~\ref{denat}B). The mid-point $[GuHCl]$ at which the protein unfolds is $\approx 2.5M$. The average radius of gyration, $\langle R_g\rangle$, of Protein L as a function of $[GuHCl]$ is in quantitative agreement with the SAXS experiments\cite{Yoo12JMB} (Fig.~\ref{denat}C). The standard deviation of $R_g$ in the protein unfolded state, $\sigma_{R_g} \approx 4$\AA, indicates that $R_g$  fluctuates between 22\AA$ \lessapprox R_g \lessapprox $30\AA. As $[GuHCl]$ is diluted from 8M to 4M, the $\langle R_g\rangle$ decreases from $\approx 26.5$\AA \ to $\approx 24.7$\AA \ almost linearly with a slope of 0.42 \AA M$^{-1}$. The experimental data\cite{Yoo12JMB} fits equally well with a horizontal line or a line with slope 0.33$\pm0.35$ \AA M$^{-1}$. The average $R_g$ of the protein conformations in the UBA basin, $\langle R_g^{UBA} \rangle$, as a function of $[GuHCl]$ show that the size of the protein decreases from $\approx 26.5$\AA \ to $\approx 22.5$\AA \ as $[GuHCl]$ is diluted from 7.5M to 0.25M.   \\
 
{\bf Denaturant-dependent FRET efficiency:} Average  FRET efficiency, $\langle E \rangle$, as a function of $[GuHCl]$ is computed from the Langevin dynamics simulations (Fig.~\ref{fret}A). In FRET experiments\cite{Sherman06PNAS, Merchant07PNAS}, the donor (AlexaFluor 488) and acceptor (AlexaFluor 594) dyes are attached near the N and C termini of Protein L. All-atom simulations\cite{Zerze15BiophysJ} have shown that the dyes have negligible effect on the size of  disordered protein structures. The $\langle E \rangle$ is calculated using
\begin{equation}\label{freteq}
\langle E \rangle = \int\limits_0^L \frac{P(R_{ee})}{1+(\frac{R_{ee}}{R_0})^6}dR_{ee},
\end{equation}
where $P(R_{ee})$ is the end-to-end distance, $R_{ee}$, probability distribution function of the protein, $L(=248$\AA$)$ is the contour length of the protein, and $R_0(=54$\AA$)$ is the Forster radius for the donor-acceptor dyes used in the experiments\cite{Sherman06PNAS, Merchant07PNAS}. 

The equilibrium $\langle E \rangle$ transitions from lower values ($<0.5$) to higher values ($\approx 0.85$) as the protein folds from an unfolded state upon $[GuHCl]$ dilution (Fig.~\ref{fret}A). The large standard-deviation, $\sigma_E \approx 0.3$, for $[GuHCl] > 2.5M$ (Fig.~S3) indicates that the protein in the UBA basin samples conformations with large size fluctuations in agreement with the $R_g$ data (Fig.~\ref{denat}C). The average FRET effeciency computed for the UBA ensemble, $\langle E^{UBA} \rangle$, to study whether the protein collapses in the early stages of folding is in quantitative agreement with the experiments of Eaton {\it et al.}\cite{Merchant07PNAS}, where as they are in disagreement with the experiments of Haran {\it et al.}\cite{Sherman06PNAS} for the denaturant concentrations $1M \lessapprox [GuHCl] \lessapprox 3M$ (Fig.~\ref{fret}A). $\langle E^{UBA} \rangle$ gradually increases from 0.37 to 0.6 as $[GuHCl]$ is diluted from 8M to 0.25M pointing to an average decrease in the size of the protein (Fig.~\ref{fret}A).

The average FRET efficiency $\langle E^{Burst} \rangle$ is also computed from the initial 0.25 $milliseconds \ (ms)$ of the Brownian dynamics simulations performed to study the folding kinetics of Protein L. The initial $0.25 \ ms$ of the folding trajectories are used to check whether the protein decreases in size in the burst-phase of folding when $[GuHCl]$ is diluted from 7.5M to lower concentrations. The initial unfolded protein conformation to initiate the folding simulations in various $[GuHCl]$ are obtained from simulations performed at $[GuHCl] = 7.5M$. 20 independent simulations starting from different initial protein conformations are performed for each $[GuHCl]$. $\langle E^{Burst} \rangle$ computed for the early stages of folding is in quantitative agreement with the experiments of Eaton {\it et al.}\cite{Merchant07PNAS} for all $[GuHCl]$, and deviates from the values obtained from the experiments of Haran {\it et al.}\cite{Sherman06PNAS} for the $[GuHCl]$ range $1M \lessapprox [GuHCl] \lessapprox 3M$ (Fig.~\ref{fret}B). $\langle E^{Burst} \rangle$ increases from 0.38 to 0.53 as $[GuHCl]$ is diluted from 7.5M to 1.0M signifying that the protein on an average decreases in size. The standard deviation of burst-phase FRET efficiency, $\sigma_E \approx 0.3$, show that $\langle E^{Burst} \rangle$ varies between 0.2 and 0.8 indicating that the protein in the initial stages of folding samples conformations with a significant variation in size (Fig.~\ref{fret}B). The $R_g$ plot as a function of time shows that it varies in the range 15\AA $\lessapprox R_g \lessapprox$ 30\AA \ during the initial hundreds of microseconds after folding is initiated for $[GuHCl]=1M$ and $2M$ conditions (Fig.~S4). \\

{\bf FRET overestimates radius of gyration in high $[GuHCl]$:} We mimicked the FRET experiments\cite{Merchant07PNAS, Sherman06PNAS} to estimate $\langle R_g^{FRET} \rangle$ from $\langle E^{Burst} \rangle$. The Gaussian polymer chain end-to-end probability distribution is given by 
\begin{equation}\label{gauss_pree}
P(R_{ee})=4 \pi R_{ee}^2 \left ( \frac{3}{2 \pi \langle R_{ee}^2  \rangle} \right)^{3/2}  \exp \left(-\frac{3 R_{ee}^2}{2 \langle R_{ee}^2  \rangle}\right).
\end{equation}
The $P(R_{ee})$ given by Eq.~\ref{gauss_pree} is used in Eq.~\ref{freteq} to estimate the average end-to-end distance square, $\langle R_{ee}^2  \rangle$, from $\langle E \rangle$. $\langle  R_g \rangle $ is calculated using the relation\cite{polymerphysics}, $\langle  R_g \rangle =\sqrt{\langle R_{ee}^2 \rangle/6}$. $\langle R_g^{FRET} \rangle$ values estimated from $\langle E^{Burst} \rangle$ (Fig.~\ref{fret}B) using equations \ref{freteq} and \ref{gauss_pree} at different $[GuHCl]$ are in near quantitative agreement with the experimentally\cite{Merchant07PNAS} estimated values (Fig.~\ref{rg_fret}A). On diluting $[GuHCl]$ from 7.5M to 1M, $\langle R_g^{FRET} \rangle$ decreases from $\approx 30$\AA \ to $\approx 24$\AA, nearly a 6\AA \ change in the size of the protein, which is in agreement with the experiments\cite{Merchant07PNAS} of Eaton {\it et al}. However, $\langle R_g^{FRET} \rangle$ deviates from the $\langle R_g^{Burst} \rangle$ values computed directly from the protein conformations obtained from the simulation trajectories (Fig.~\ref{rg_fret}A). $\langle R_g^{Burst} \rangle$ decreases from $\approx 26.7$\AA \ to $\approx 23.8$\AA, a decrease of only $\approx 3$\AA \ upon $[GuHCl]$ dilution from 7.5M to 1M (Fig.~\ref{rg_fret}A). This shows that FRET overestimates the size of the protein especially in higher $[GuHCl]$, and this gives rise to the appearance of pronounced compaction in the dimensions of the protein in the burst-phase as $[GuHCl]$ is diluted. The standard deviation, $\sigma_{R_g} \approx 4$\AA, of $\langle R_g^{Burst} \rangle$ shows that at all $[GuHCl]$, the protein samples conformations with $R_g$ varying from $\approx 22$\AA \ to $\approx 28$\AA \ (Fig.~\ref{rg_fret}A), and the 3\AA \ decrease in $\langle R_g^{Burst} \rangle$ upon $[GuHCl]$ dilution is within the $\sigma_{R_g}$. The deviation between $\langle R_g^{FRET} \rangle$ and $\langle R_g^{Burst} \rangle$ at high $[GuHCl]$ was also emphasised in the work of O'Brien {\it et al.}\cite{OBrien09JCP} and the reasons for the deviation are attributed to the use of Gaussian chain $P(R_{ee})$ to extract $\langle R_g^{Burst} \rangle$. The problems associated with the use of Gaussian chain $P(R_{ee})$ to extract information about protein dimensions is also highlighted in previous other studies\cite{Sinha07JMB,Laurence05PNAS,Goldenberg03JMB,Zhou02JPCB}. The deviation between $\langle R_g^{FRET} \rangle$ and $\langle R_g^{Burst} \rangle$ increases with $[GuHCl]$, and the reasons for the discrepancy can be understood using the relation $\langle R_{ee}^2 \rangle = \langle R_{ee} \rangle^2 + \sigma_{R_{ee}}^2$, where $\sigma_{R_{ee}}$ is the standard deviation in $R_{ee}$.

$\langle R_{ee}^{FRET} \rangle$ and $\sigma_{R_{ee}}^{FRET}$ estimated from the Gaussian polymer chain $P(R_{ee})$ to compute $\langle R_g^{FRET} \rangle$ deviate from the values $\langle R_g^{Burst} \rangle$ computed directly from the initial $0.25 ms$ of the Protein L folding trajectories (Fig.~\ref{rg_fret}B and C). At high $[GuHCl] (= 7.5M)$, $\langle R_{ee}^{FRET} \rangle$ and $\sigma_{R_{ee}}^{FRET}$ estimated from the Gaussian chain $P(R_{ee})$ are 67.3\AA \ and 28.3\AA, respectively, which deviate from the $\langle R_{ee}^{Burst} \rangle$ and $\sigma_{R_{ee}}^{Burst}$ values 62.7\AA \ and 20.5\AA \ respectively (Fig.~\ref{rg_fret}B), computed directly from the simulations. As a result FRET overestimates \\ $\langle R_g^{FRET} \rangle \left(=\sqrt{\langle (R_{ee}^{FRET})^2 \rangle/6}=\sqrt{[\langle R_{ee}^{FRET} \rangle^2+(\sigma_{R_{ee}}^{FRET})^2]/6}\right)$ compared to $\langle R_g^{Burst} \rangle$  in high $[GuHCl]$ (Fig.~\ref{rg_fret}A). As $[GuHCl]$ decreases, the deviation between $\langle R_g^{FRET} \rangle$ and  $\langle R_g^{Burst} \rangle$ decreases (Fig.~S5). In low $[GuHCl](=1.0M)$, the  $\langle R_{ee}^{FRET} \rangle$ values computed from the Gaussian chain $P(R_{ee})$, and $\langle R_{ee}^{Burst} \rangle$ computed from simulations are in good agreement, where as the $\sigma_{R_{ee}}^{FRET}$ values deviate from $\sigma_{R_{ee}}^{Burst}$ (Fig.~\ref{rg_fret}C). Due to this the deviation between $ \langle R_g^{Burst} \rangle$ and $\langle R_g^{FRET} \rangle$ is small in low $[GuHCl]$, and  increases with $[GuHCl]$ (Fig.~\ref{rg_fret} and S5).

To conclude, during the burst-phase of folding, $\langle R_g^{Burst} \rangle$ for Protien L decreases from  $\approx 26.7(\pm4)$\AA \ to $\approx 23.8(\pm4)$\AA, a decrease of $\approx 3$\AA \ upon $[GuHCl]$ dilution from 7.5M to 1M (Fig.~\ref{rg_fret}A). However, FRET overestimates the size of the protein in high $[GuHCl](\approx 7.5M)$ due to the application of the Gaussian polymer chain $P(R_{ee})$ to estimate $\langle R_g^{FRET} \rangle$. During the burst-phase $\langle R_g^{FRET} \rangle$ estimated from FRET decreases from $\approx 30$\AA \ to $\approx 24$\AA \ upon $[GuHCl]$ dilution from 7.5M to 1M (Fig.~\ref{rg_fret}A). Due to the $\approx$6\AA \ decrease in $\langle R_g^{FRET} \rangle$, FRET experiments\cite{Merchant07PNAS, Sherman06PNAS} suggest pronounced compaction in the protein size during the burst-phase of folding. \\

{\bf Disagreement between FRET and SAXS experiments on Protein L compaction in burst-phase folding:} In simulations, the average radius of gyration in the burst-phase folding, $\langle R_g^{Burst} \rangle$, decreased by  $\approx3$\AA \ when $[GuHCl]$ is diluted from 7.5M to 1M. The $\approx3$\AA \ decrease in $\langle R_g^{Burst} \rangle$ is close to statistical uncertainties of the $R_g$ data obtained from SAXS experiments\cite{Yoo12JMB} for Protein L. The $R_g$ data from SAXS experiments for $3M \lessapprox  [GuHCl] \lessapprox 7M$ can fit a horizontal line or a line with slope 0.33$\pm0.35$ \AA M$^{-1}$ equally well\cite{Yoo12JMB}. Using this slope to compute the $R_g$ change upon $[GuHCl]$ dilution from $\approx $7.5M to 1M gives a $R_g$ decrease of $2.1\pm2.3$\AA. The SAXS experiments\cite{Yoo12JMB} report that $R_g$ of Protein L in the burst phase upon $[GuHCl]$ dilution to 1.3M and 0.67M are $\approx 23.5\pm2.1$\AA \ and 24.9$\pm1.12$\AA, respectively which are statistically not different from the value $23.7\pm0.4$ at $[GuHCl]=4.0$M. In the simulations, $\langle R_g^{Burst} \rangle$ at $[GuHCl]$ = 4M and 1M are $25.1\pm4.0$\AA \ and $23.8\pm3.9$\AA, respectively, a difference of $\approx 1.3$\AA, which is well within $\sigma_{R_g}\approx$ 4\AA. This analysis similar to the SAXS analysis leads to the conclusion of minimal Protein L compaction within statistical uncertainties on $[GuHCl]$ dilution in agreement with the SAXS experiments\cite{Yoo12JMB}, and disagreement with the FRET experiments\cite{Sherman06PNAS, Merchant07PNAS}.

Recent FRET experiments\cite{Watkins15PNAS} on polyethylene glycol (PEG) showed that FRET efficiency decreased as $[GuHCl]$ is increased when hydrophilic PEG is unlikely to expand on increasing $[GuHCl]$. This led to questions about the interpretation of the FRET data to study protein collapse in low $[GuHCl]$. We find that the computed variation in $\langle E \rangle$ and $\langle R_g \rangle$ as a function of $[GuHCl]$ for Protein L to be in quantitative agreement with at least one of the FRET experiments\cite{Merchant07PNAS} (Fig.~\ref{fret}) and also in agreement with SAXS experiments\cite{Yoo12JMB} within the statistical uncertainties (Fig.~\ref{denat}C and ~\ref{rg_fret}A). The results points to the use of Gaussian polymer chain statistics to extract $R_g$ from FRET efficiency data to be the cause for the discrepancy between the SAXS and FRET experiments in estimating $R_g$. \\

{\bf The coil-globule transition in Protein L is concomitant with the folding transition}: In polymers the ratio of the radius of gyration to the hydrodynamic radius, $R_g/R_h$, can point to the coil-globule collapse transition. The  $R_g/R_h$ ratio for a polymer in a good solvent\cite{Oono83JCP} is $\approx 1.56$, where as the ratio in a poor solvent\cite{polymerphysics} is $\approx 0.77$. We used the Kirkwood-Riseman approximation\cite{Kirkwood48JCP} to compute the hydrodynamic radius  of the protein, which is given by $R_h = (1/2N^2)\sum\limits_{i \neq j} 1/\left| \vec{r_i}-\vec{r_j} \right|$, where $N$ is the number of beads in the coarse-grained protein, $\vec{r_i}$ and $\vec{r_j}$ are the position vectors of beads $i$ and $j$. The $ R_g / R_h $ ratio for the burst-phase folding decreases from $\approx 1.31$ to $\approx 1.28$ as $[GuHCl]$ is diluted from 7.5M to 1M indicating that this is not a coil-globule transition observed in polymers (Fig.~\ref{ratio}). The single domain proteins which are finite in size compared to polymers are predicted to have a near overlap of the collapse and folding transition temperatures\cite{Camacho93PNAS,Li04PRL}. In agreement, the equilibrium ratio of $ R_g/ R_h $ decreases from $\approx 1.3$ to $\approx 0.98$ as the folding transition occurs (Fig.~\ref{ratio}). The $ R_g/ R_h $ ratio does not approach 0.77 as the protein folds because the Kirkwood-Riseman approximation\cite{Kirkwood48JCP} used to compute $R_h$ does not hold for the protein in the folded state as it assumes all the beads are equally bathed by the solvent. 
The absence of coil-globule transition in the burst-phase of protein L folding does not imply that the collapse transition or significant protein compaction is universally absent in the burst-phase folding of all single domain proteins. Both FRET and SAXS experiments agree that the protein Monellin\cite{Reddy15Biochemistry,Jha09PNAS,Kimura05PNAS} shows compaction during the burst-phase of folding. Although both the experimental techniques observe compaction in the case of Cytochrome {\it c}\cite{Kathuria14JMB,Lapidus07BiophysicsJ,Hagen00JMB,Shastry98ACS}, the FRET experiments show that this compaction, a sub-$100 \mu s$ event, is barrier limited and it is due to the formation of marginally stable partially folded structures\cite{Kathuria14JMB}. Experiments\cite{Hofmann12PNAS} show that for the  protein CyclophilinA, the $R_g/R_h$ ratio decreases from a value between 1.1-1.2 to a value between 0.9-1.0 as $[GuHCl]$ is diluted from 8M to 0M indicating a coil-globule transition in the burst-phase of folding. \\
 
{\bf Transition State Ensemble (TSE)}: The transition state ensemble of Protein L at the melting temperature, $T_M$, is identified using the $P_{fold}$ analysis\cite{Du98JCP} (see SI for details). 12 out of 108 putative transition state structures (TSE) which satisfy the condition, $0.4< P_{fold} <0.6$ are labeled as TSE (Fig.~S6). The transition state structures (TSE) are globular, extensive and homogenous, with most of the secondary and tertiary contacts formed (Fig.~\ref{trans}). The $\Psi$-analysis experiments\cite{Yoo12JMB2} predict that TSE contains all the four $\beta$-sheet strands ($\beta_1 - \beta_4$). The TSE from simulations show that both the $N$ and $C$-termini hairpins $\beta_1\beta_2$ and $\beta_3\beta_4$, and the contacts between the strands $\beta_1\beta_4$ are present in the structures in agreement with the $\Psi$-analysis experiments\cite{Yoo12JMB2} (Fig.~\ref{trans}B). 

The  $\Psi$-analysis experiments on two residue pairs, K28-E32 and A35-T39, present in the helical region of the protein gave $\Psi$-values 0.26 and $\ll 0$, respectively, indicating that contacts between these pairs of residues is largely absent, and concluded that helix $\alpha_1$ is mostly not present in the TSE\cite{Yoo12JMB2}.  The contact map of the TSE obtained from the simulations show that the side chains of the residue pairs K28-E32 and A35-T39 form contacts with a probability of 0.41 and 0.08, respectively. The simulations further indicate that a cluster of residues between S31 and A37 present approximately at the center of the helix containing 3 Ala residues (A33, A35 and A37) can form stable contacts in the TSE (Fig.~\ref{trans}B).

The $\Psi$-analysis experiments\cite{Pandit06JMB, Baxa08JMB} predict a relationship between the relative contact order, RCO (see methods), of the native protein topology and TSE, $RCO^{TSE}  \approx 0.7 RCO^{Native}$, which shows the extent of long-range contacts present in the TSE compared to the native-state. The TSE structures extracted from the simulations show $RCO^{TSE}/RCO^{Native}=0.77$, which is in reasonable agreement with the value of 0.75 estimated from $\Psi$-analysis experiments\cite{Yoo12JMB2}. The simulations using the coarse-grained protein model support the basic topology of the TSE structures predicted by the $\Psi$-analysis experiments.

The folding simulations of only the C-terminal hairpin ($\beta_3\beta_4$) using atomistic models predicted the presence of non-native contacts, a 2 amino acid register shift, in the TSE\cite{Yoo12JMB2}. We do not observe this 2 amino acid register shift in the C-terminal hairpin because the SOP-SC model includes only native-interactions. The predicted TSE is only in partial agreement with the $\Phi$-analysis experiments\cite{Kim00JMB,Kim98JMB,Scalley97Biochem,Gu97JMB} which predicted a polarised structure with only $\beta_1\beta_2$ hairpin. The results support the hypothesis that folding pathways and TSE of single domain proteins are influenced by the topology of the folded structure in agreement with the experiments\cite{Baxa15PNAS}. \\
 
{\bf Concluding Remarks:}
In summary, we have studied Protein L folding in the presence of the denaturant Guanidine Hydrochloride using the SOP-SC coarse-grained model and molecular dynamics simulations. The effect of $[GuHCl]$ on the protein is taken into account using the molecular transfer model\cite{OBrien08PNAS,Liu12JPCB}. The study mainly focussed on whether there is a coil-globule collapse transition in the burst-phase of folding after the denaturant concentration is diluted to lower values. The main findings of this study is the coil-globule transition in Protein L is concomitant with the folding transition. It is not observed during the burst phase. The FRET experiments overestimate the $R_g$ of the protein at high $[GuHCl]$ concentrations owing to the use of the Gaussian polymer chain end-to-end distribution function to extract $R_g$ from FRET efficiency. As a result, in the burst-phase of folding, FRET observes pronounced compaction in the size of the protein  as $[GuHCl]$ is diluted. The actual decrease in the size of the protein ($\approx 3$\AA) observed during the burst-phase is close to statistical uncertainties of the $R_g$ data measured from SAXS experiments\cite{Yoo12JMB}, and these experiments conclude that there is no collapse leading to a discrepancy with the FRET experiments.

It is highly desirable to formulate a method to accurately extract the distances between the donor and acceptor dyes used in the FRET experiments. However, it is a non-trivial inverse problem as we seek to accurately extract a probability distribution of the distances between the dyes from the average FRET efficiency measured in experiments, especially in cases like Protein L where the compaction in protein dimensions is small on denaturant dilution. Previous studies\cite{OBrien09JCP} have shown that even other polymer models such as the self-avoiding chain or the worm-like-chain model are also not very accurate quantitatively to predict the small subtle changes in the protein dimensions.  

The results presented in this manuscript clearly point out the aspects of the Gaussian chain model, which leads to over estimating the size of the protein when used to analyse the FRET data. The results show that the Gaussian chain model  fails in accurately capturing the width of the protein end-to-end probability distribution, which is essential to compute the radius of gyration. For any method to be quantitatively accurate it should capture the peak position as well as the width of the probability distribution accurately, and this is a challenging task because we need to estimate probability distribution from an average value, and also the method should be reliable enough to work on proteins with different amino acid composition and native folds. 

To check the accuracy of the distance between the dyes extracted from the FRET efficiency data using the Gaussian polymer model assumption, a self-consistency check can be performed to see if the assumption is valid or not for the protein under study\cite{OBrien09JCP}. If the dyes are attached at locations $i$ and $j$ in the protein, and $\langle R_{ij}^2 \rangle$ is the average distance square extracted from FRET efficiency, and similarly if $\langle R_{kl}^2 \rangle$ is the average distance square extracted from FRET efficiency with dyes at positions $k$ and $l$, then the relation
                     $\langle R_{ij}^2 \rangle / \langle R_{kl}^2 \rangle = {|j-i|/l-k|}$  should hold if the protein behaves as a Gaussian chain. If the relation is not satisfied, then one should be cautious in quantitatively inferring results about the protein dimensions assuming that the protein in the unfolded state behaves as a Gaussian chain.

The magnitude of protein compaction in the burst-phase folding of single domain proteins upon denaturant dilution is not uniform, and it should depend on protein length, sequence and composition of amino acids. For example SAXS experiments on Protein L\cite{Yoo12JMB} and Ubiquitin\cite{Jacob04JMB} infer no compaction in the protein dimensions in the burst-phase, while experiments on Cytochrome {\it c}\cite{Kathuria14JMB} and Monellin\cite{Kimura05PNAS} observe compaction. The key features in the single domain proteins responsible for compaction in protein dimensions on denaturant dilution needs to be identified. In addition to temperature and denaturants, force can also be used to unfold proteins and study protein folding. Experiments\cite{Fernandez04Science} and simulations\cite{Best05Science, Hyeon09PNAS} show that a protein unfolded by force when allowed to refold in the presence of lower quenching forces undergoes a rapid compaction in the initial stages of folding. This compaction of the protein is driven by entropy because the protein in the stretched state is in a low entropic state and upon force quench undergoes rapid compaction in the first stage of folding until entropy is maximised\cite{Hyeon09PNAS}. The extent of protein compaction in the initial stages of folding also depends on the experimental probes used to study protein folding.

The transition state structures inferred from the $P_{fold}$ analysis are globular and extensive with both the C and N-termini hairpins $\beta_1\beta_2$ and $\beta_3\beta_4$, and interactions between the strands $\beta_1\beta_4$. These results are in agreement with the $\Psi$-analysis experiments\cite{Yoo12JMB2} and support the hypothesis that for single domain globular proteins the transition state structures depend on the protein native-state topology and are not stabilised by local interactions alone.
 
{\bf Acknowledgement:} 
GR acknowledges startup grant from Indian Institute of Science-Bangalore, and funding from Nano mission, Department of Science and Technology, India. Hiranmay Maity acknowledges research fellowship from Indian Institute of Science-Bangalore.

\bigskip

{\bf Supporting Information:} Description of the simulation methods; Table S1; Figures S1-S6.  This material is available free of charge via the Internet at http://pubs.acs.org

\newpage

\bibliography{Proteins12}
\bibliographystyle{unsrt}

\newpage

\begin{figure} 
\includegraphics[width=3in]{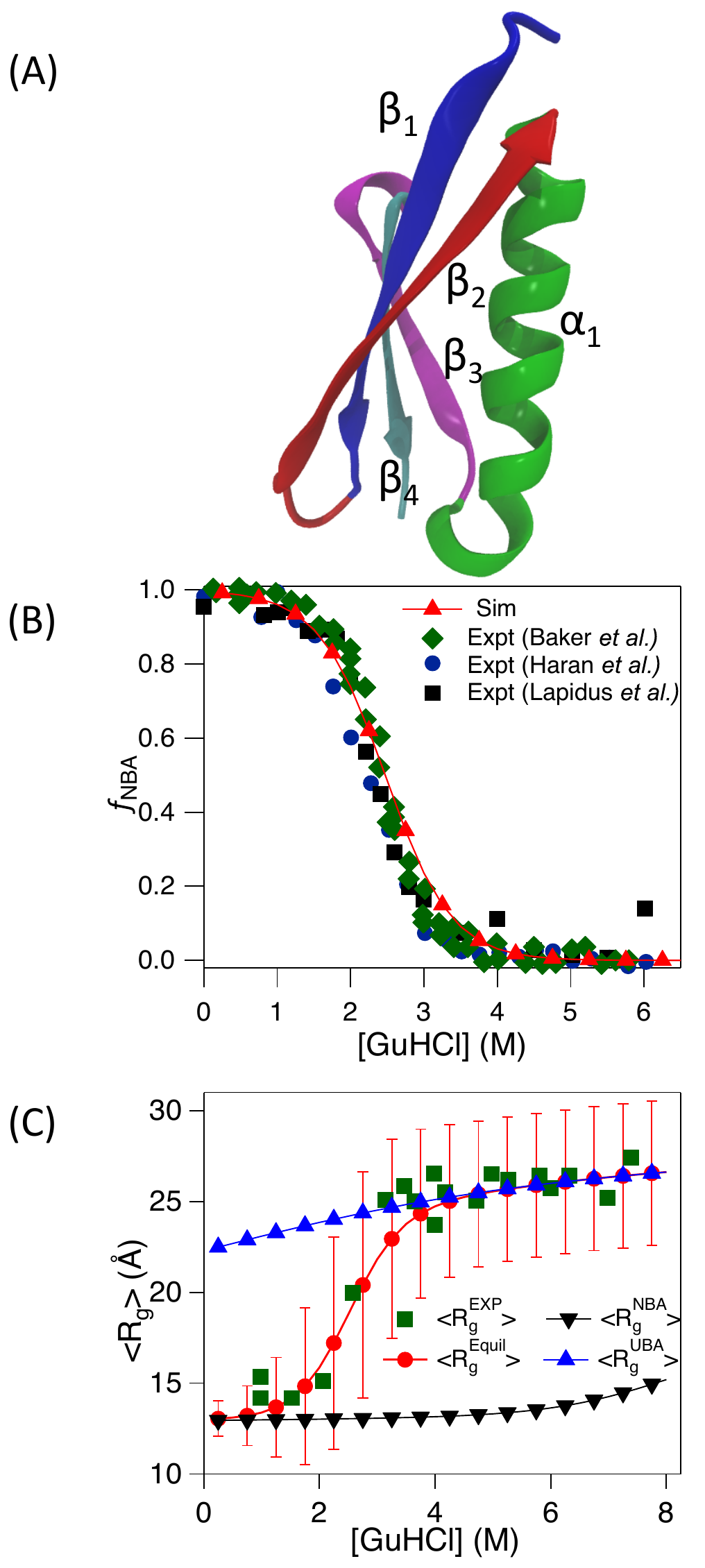}
\caption{(A) Crystal structure of Protein L (PDB ID: 1HZ6). The $\alpha$-helix is in green ($\alpha_1$), and the four $\beta$-strands are in blue ($\beta_1$), red ($\beta_2$), magenta ($\beta_3$), and cyan ($\beta_4$). (B) The fraction of the protein in the native basin of attraction, $f_{NBA}$, as a function of $[GuHCl]$. Data in red triangles is from simulations. Data in blue circles, black squares and green diamonds are from the experiments of Haran {\it et al.}\cite{Sherman08Biophys}, Lapidus {\it et al}\cite{Waldauer08HFSP} and Baker {\it et al.}\cite{Kim00JMB} respectively. (C) The radius of gyration, $R_g$ as a function of $[GuHCl]$. Data in red circles and green squares is from simulations and experiments\cite{Yoo12JMB}, respectively. $\langle R_g \rangle$ of UBA and NBA basins computed from simulations are shown in blue triangles and black inverted triangles, respectively.}\label{denat}
\end{figure}

\begin{figure} 
\includegraphics[width=3.5in]{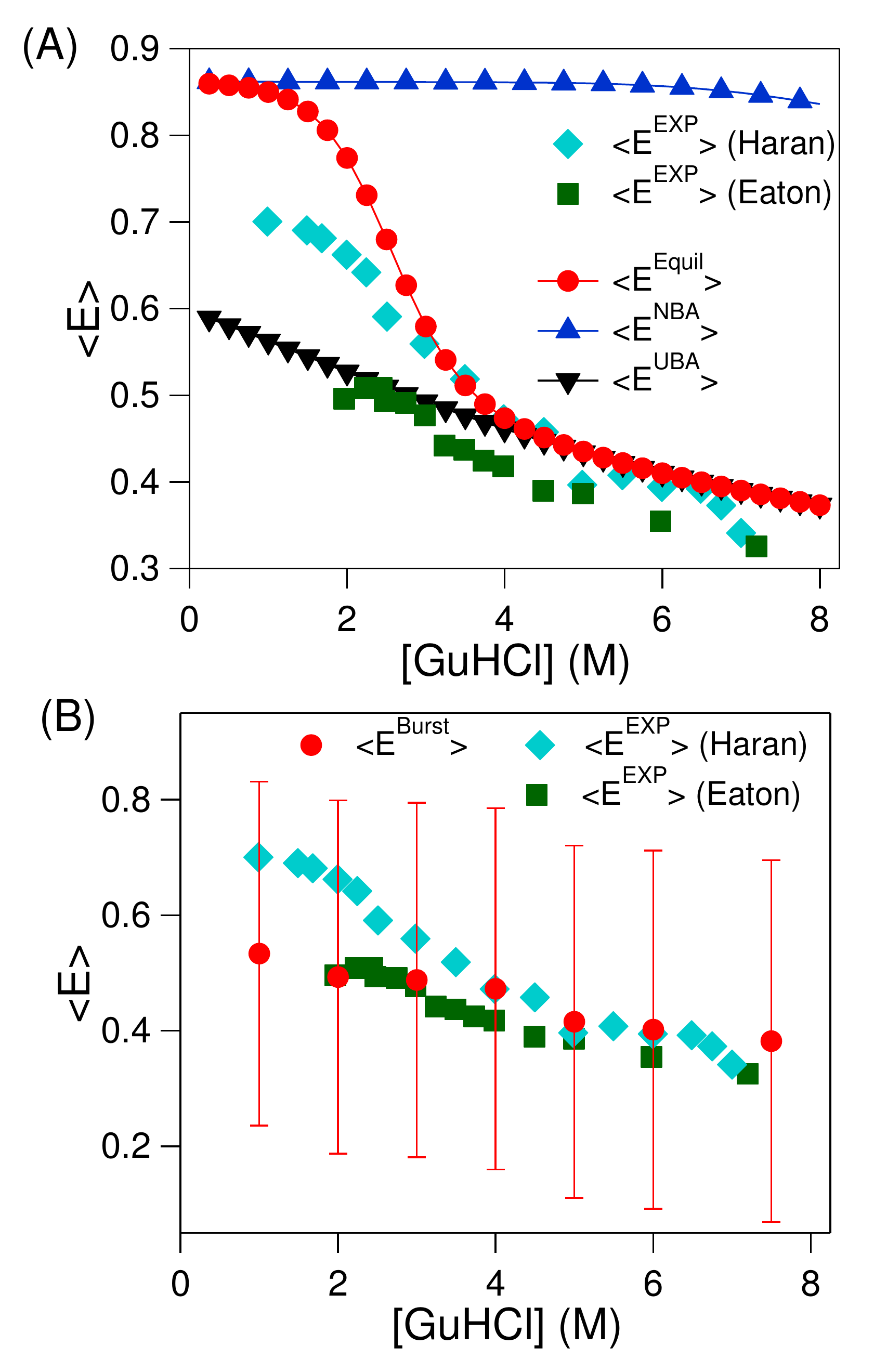}
\caption{(A) Equilibrium FRET efficiency, $\langle E^{Equil} \rangle$, as a function of $[GuHCl]$ is in red circles. Experimental data are shown in green squares\cite{Merchant07PNAS} and cyan diamonds\cite{Sherman06PNAS}. $\langle E \rangle$ for the protein conformations in the NBA and UBA basins are shown in blue triangles and black inverted triangles, respectively. (B) $\langle E^{Burst} \rangle$ shown in red circles is computed from the initial $0.25 \ ms$ of Protein L Brownian dynamics folding trajectories at $T=357.7K$ in various $[GuHCl]$ . Data in green squares and cyan diamonds is the same as in (A).}\label{fret}
\end{figure}

\begin{figure} 
\includegraphics[width=3.5in]{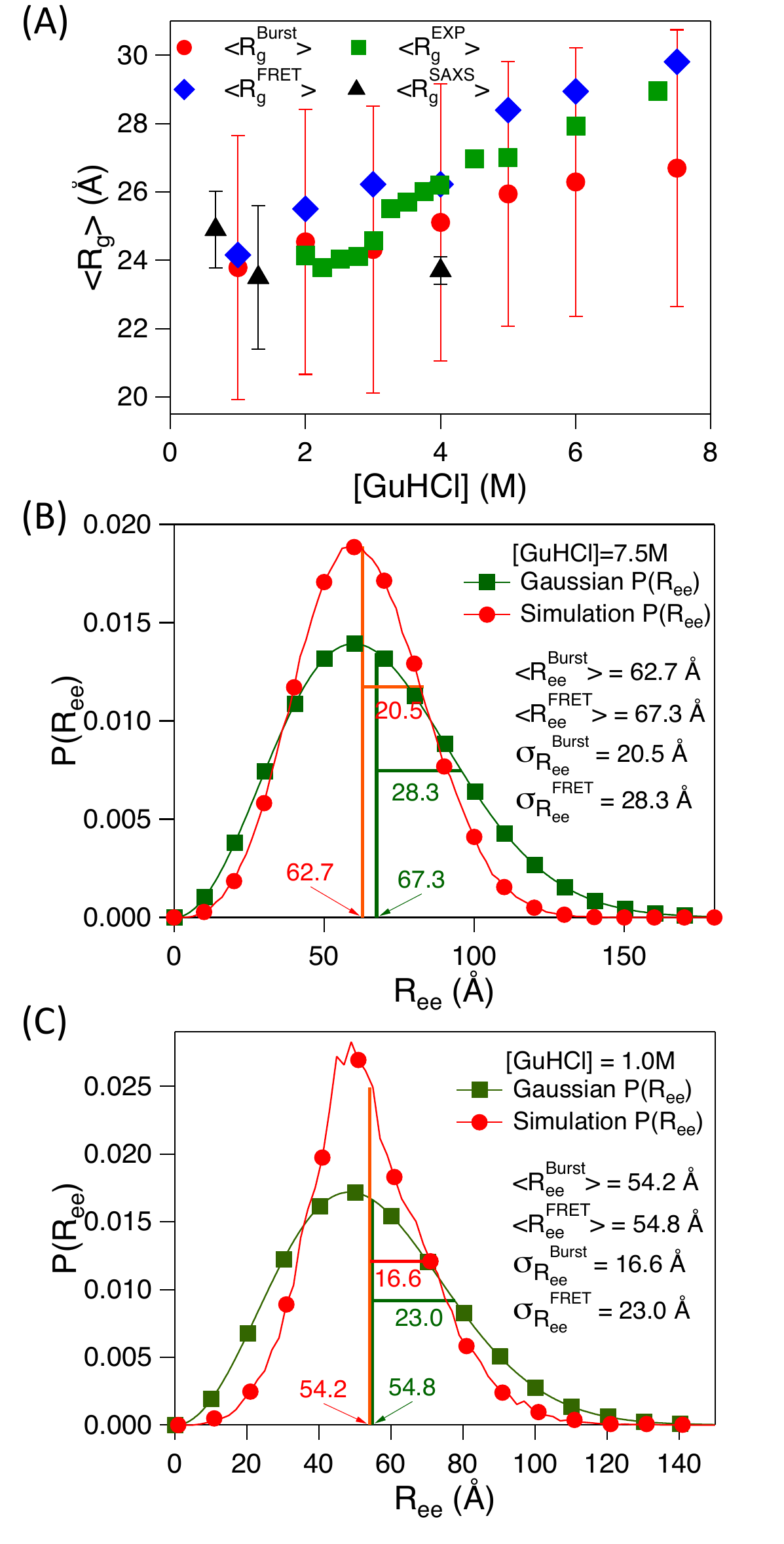}
\caption{(A) $\langle R_g^{FRET} \rangle$ estimated from $\langle E^{Burst} \rangle$ is in blue diamonds. $\langle R_g^{Burst} \rangle$ computed from the initial $0.25ms$ of the Protein L folding trajectories is in red circles.  Data in green squares and black triangles is from FRET\cite{Merchant07PNAS} and SAXS\cite{Yoo12JMB} experiments, respectively. (B) The end-to-end distance, $R_{ee}$, probability distribution function $P(R_{ee})$ during the burst phase (initial $0.25ms$) of protein L folding at $T=357.7K$ and $[GuHCl]=7.5M$ is in red circles. $P(R_{ee})$ estimated from $\langle E^{Burst} \rangle$ and Guassian polymer chain statistics in green squares. (C) same as in (B) except that $[GuHCl]=1.0M$.}\label{rg_fret}
\end{figure}

\begin{figure} 
\includegraphics[width=4in]{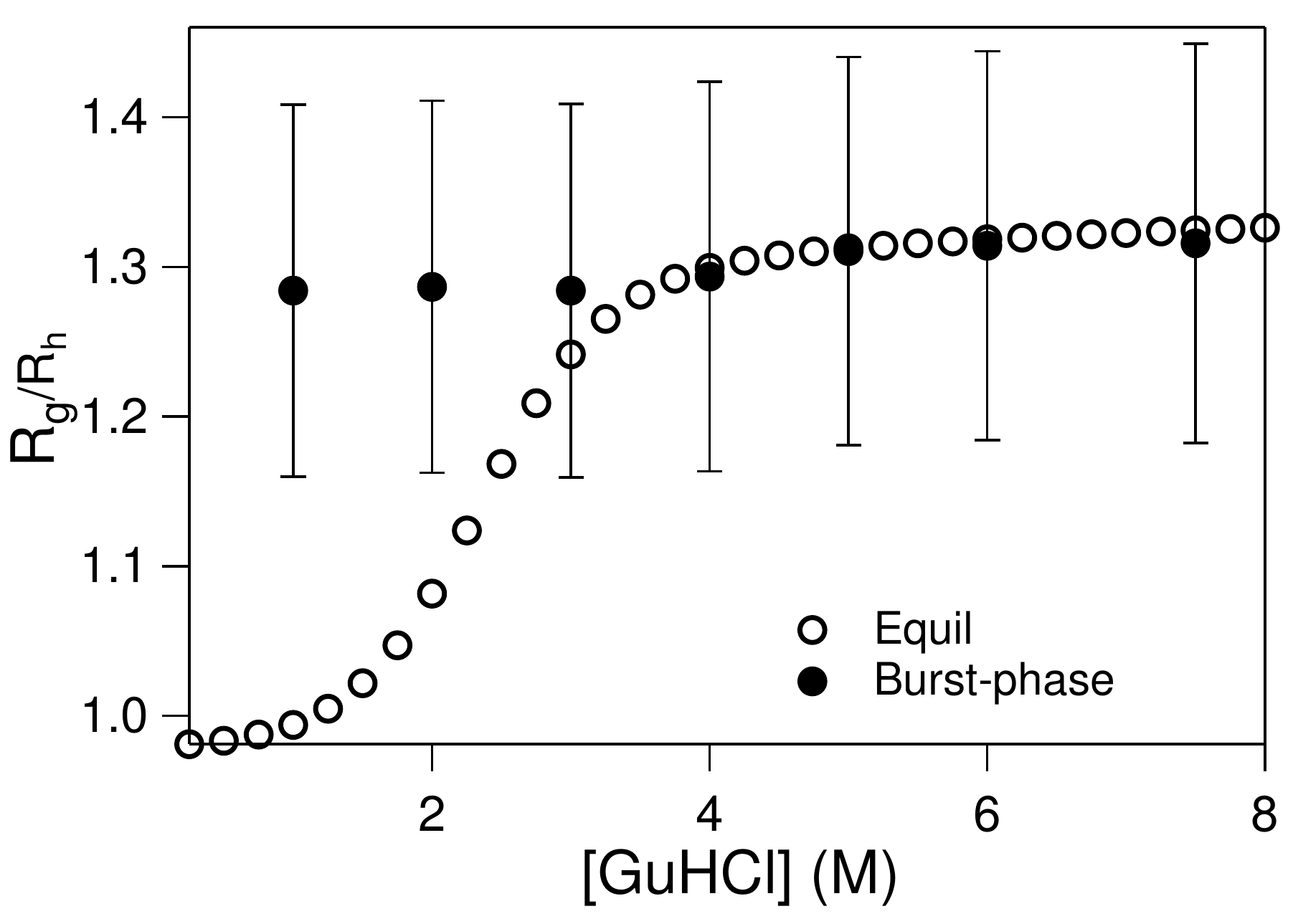}
\caption{The ratio of the radius of gyration to the hydrodynamic radius, $ R_g / R_h $, for the burst phase (solid circles) and equilibrium (empty circles) conditions.}\label{ratio}
\end{figure}

\begin{figure} 
\includegraphics[width=4in]{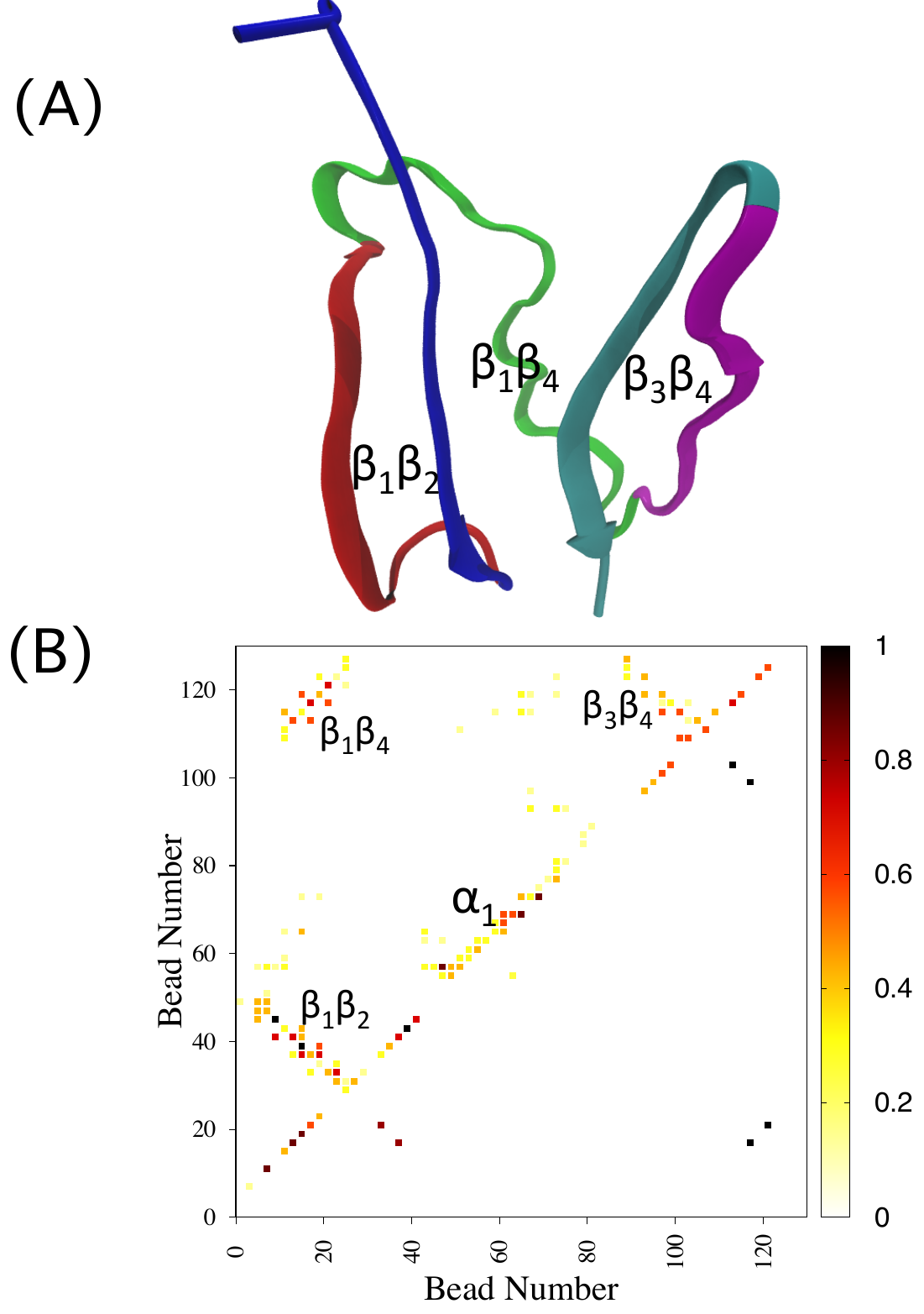}
\caption{(A) Representative transition state structure from the transition state ensemble obtained using the $P_{fold}$ analysis. (B). The contact map of the the transition state ensemble is shown in the upper half of the diagonal. The experimental\cite{Yoo12JMB2} $\Psi$-values for the transition state structure are show in the lower half of the diagonal. The $\Psi$-values for the $\alpha$-helix residue pairs K28-E32 and A35-T39 from experiments\cite{Yoo12JMB2} are 0.26 and $\ll 0$, respectively. The $\Psi$-value used for the residue pair A35-T39 is 0 in the plot. The small $\Psi$-values for $\alpha$-helix are not visible in the plot.}\label{trans}
\end{figure}

\begin{figure}
\includegraphics[width=3.25in]{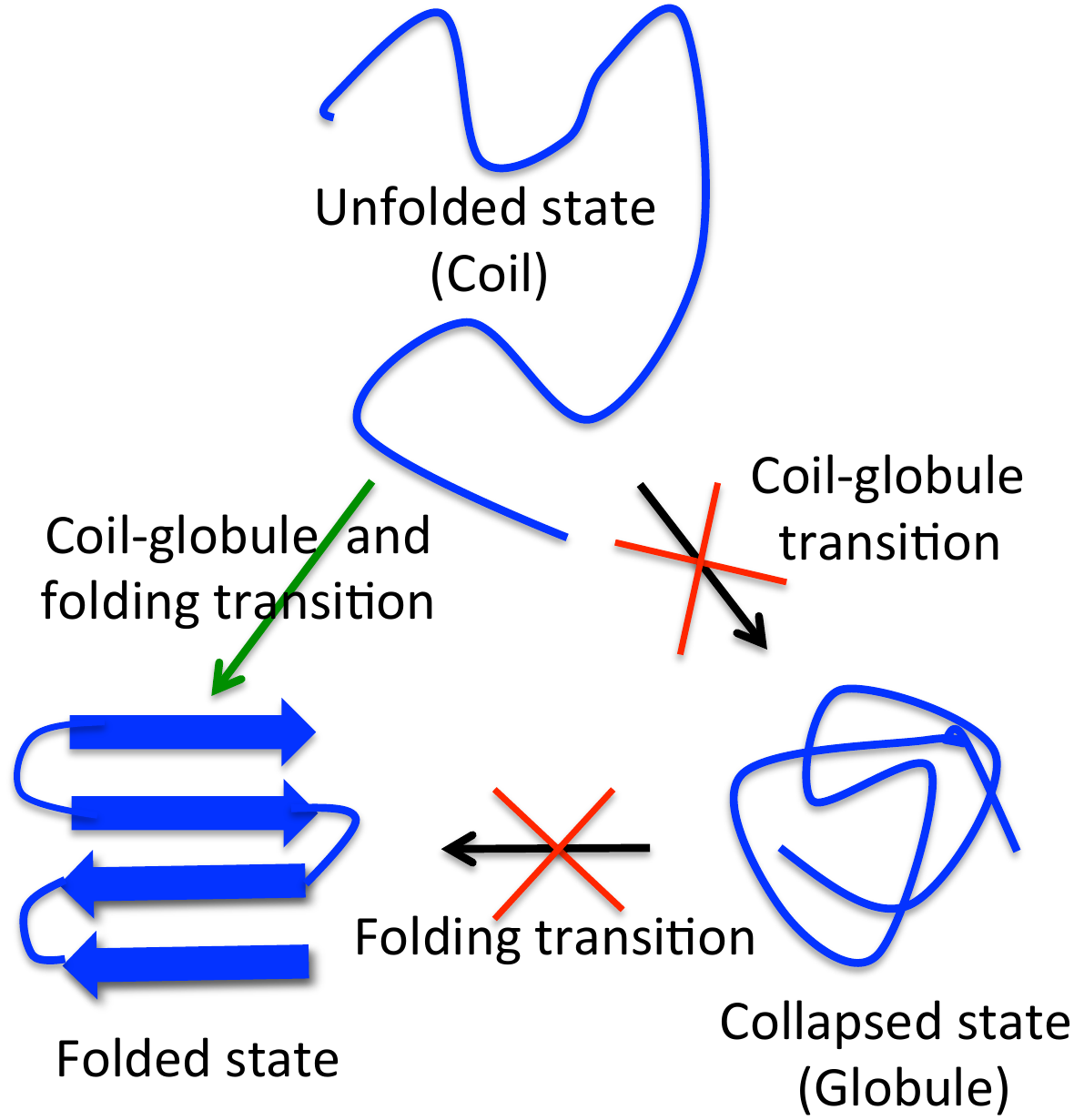}
\caption{Table of Contents (TOC) figure}
\label{toc}
\end{figure}

\end{document}


\setcounter{figure}{0}
\setcounter{table}{0}
\setcounter{equation}{0}
\renewcommand{\thefigure}{S\arabic{figure}}
\renewcommand{\thetable}{S\arabic{table}}
\renewcommand{\theequation}{S\arabic{equation}}
						
\title{Supporting Information for ``Folding of Protein L with implications for collapse in the denatured state ensemble"}
\author{Hiranmay Maity}
\author{Govardhan Reddy}
\email{greddy@sscu.iisc.ernet.in}
\affiliation{Solid State and Structural Chemistry Unit, Indian Institute of Science, Bangalore, Karnataka, India 560012}

\maketitle
\newpage

\textbf{SI Text}

\bigskip

{\bf Self Organized Polymer-Side Chain (SOP-SC) model for Protein L:} 

We used the SOP-SC (self-organized polymer-side chain) model\cite{Hyeon06Structure,Liu11PNAS} in which each amino acid residue is represented by two beads. One bead is at the $C_\alpha$ position representing the backbone atoms, and the other bead is at the center of mass of the side chain representing the side chain atoms. The number of residues in Protein L, $N_{res}=64$. The effective energy of a protein conformation in the SOP-SC model is a sum of bonded and non-bonded interactions. The bonded interactions, $E_B$, are present between a pair of  connected beads. The non-bonded interactions are a sum of native, $E_{NB}^{N}$, and non-native, $E_{NB}^{NN}$, interactions.  The native interactions for protein L are identified using the crystal structure\cite{Neill01ActaCry} (Protein Data Bank ID: 1HZ6) (Fig 1A), and they are present between a pair of beads separated by at least 3 bonds, and if the distance between them in the crystal structure is less than $R_c$ (Table~S1). 

The coarse-grained force-field in the SOP-SC model for a protein conformation represented by the co-ordinates, $\{\bf r\}$, in the absence of denaturants, $[C]=0$, is 
\begin{equation}\label{pot}
E_{CG}(\{{\bf r}\},0)  = E_{B} + E_{NB}^{N} +  E_{NB}^{NN}. 
\end{equation}
  
The bonded interaction energy, $E_B$, for all pairs of bonded beads is modelled by finite extensible nonlinear elastic (FENE) potential,
\begin{equation}
E_B = -\sum _{i=1}^{N_B} \frac{k}{2}R_{0}^{2}log\Big(1 - \frac{(r_{i}-r_{cry,i})^2}{R_0^2}\Big),
\label{Eq. bonded_energy}
\end{equation}
where $N_B(=127)$ is the total number of pairs of bonds in the SOP-SC model of the protein. The values of $k$ and $R_0$  are listed in table S1.
Non-bonded native interaction energy, $E_{NB}^N$, is modelled by Lennard-Jones type of potential energy and is given by
\begin{equation}
\begin{split}
E_{NB}^{N}& = \sum _{i=1}^{N_{N}^{bb}} \epsilon _h^{bb}\big[ \big( \frac{r_{cry,i}}{r_i}\big)^{12} -2\big(\frac{r_{cry,i}}{r_i}\big)^6 \big ] + \sum _{i=1}^{N_{N}^{bs}} \epsilon _h^{bs}\big[ \big( \frac{r_{cry,i}}{r_i}\big)^{12} -2\big(\frac{r_{cry,i}}{r_i}\big)^6 \big ]  \\
& + \sum _{i=1}^{N_{N}^{ss}} 0.5(0.7-\epsilon _i^{ss})300k_B\big[ \big( \frac{r_{cry,i}}{r_i}\big)^{12} -2\big(\frac{r_{cry,i}}{r_i}\big)^6 \big ],
\end{split}
\label{Eq. non_bonded_native_energy}
\end{equation}
where $N_{N}^{bb}$, $N_{N}^{bs}$ and $N_{N}^{ss}$ denote the number of native contact pairs between backbone-backbone, backbone - side chain and side chain - side chain, respectively. The values of $N_{N}^{bb}$, $N_{N}^{bs}$ and $N_{N}^{ss}$ are 172, 432 and 173, respectively. $k_B$ is the Boltzmann constant, $r_i$ denotes the distance between $i^{th}$ pair of beads, and $r_{cry,i}$ denotes the corresponding distance in the crystal structure. $\epsilon _{h}^{bb}$, $\epsilon _{h}^{bs}$ and $\epsilon _{i}^{ss}$ denote the strength of backbone - backbone, backbone - side chain and side chain - side chain interactions, respectively (Table~S1). The values of $\epsilon _{i}^{ss}$ are taken from the Betancourt-Thirumalai statistical potential \cite{Betancourt99ProtSci}.

The non-native interactions, $E_{NB}^{NN}$, are purely repulsive interactions and are given by
\begin{equation}
\begin{split}
E_{NB}^{NN} & = \sum_{i=1}^{N_{NN}} \epsilon_{l} \left(\frac{\sigma_{i}}{r_{i}} \right)^{6} + \sum_{i=1}^{N_{ang}^{bb}} \epsilon_{l} \left(\frac{\sigma^{bb}}{r_{i}} \right)^{6} + \sum_{i=1}^{N_{ang}^{bs}} \epsilon_{l} \left(\frac{\sigma_{i}^{bs}}{r_{i}} \right)^{6}
\end{split}
\end{equation}
where  $N_{NN}$(= 6973) is the total number of non-native interactions, $N_{ang}^{bb}$(= 62) is the number of pairs of backbone beads separated by 2 bonds in the SOP-SC model, and  $N_{ang}^{bs}$(= 126) is the number of pairs of backbone and side chain beads separated by 2 bonds in the SOP-SC model. $\sigma^{bb}$ is the diameter of the backbone beads, and $\sigma^{bs}_{i}(= f[\sigma^{bb}+\sigma^{sc}_{i}]/2.0)$ is the sum of the radii of the backbone and the side chain in the  $i^{th}$ pair of angular interactions scaled by a factor $f=0.9$. Values of the side chain radii are given in Table S2 in Ref.\cite{Reddy12PNAS}

The values of the parameters used in the energy function (Table~S1) are identical to the values previously used to successfully study the folding properties of the proteins GFP\cite{Reddy12PNAS} and Ubiquitin\cite{Reddy15JPCB}. We have used the same force-field to study the properties of different proteins, and as a result this force-field satisfies the criterion of a transferable force-field.

\bigskip

{\bf Molecular Transfer Model:}  To simulate Protein L folding thermodynamics and kinetics in the presence of Guanidine Hydrochloride we used the Molecular Transfer Model (MTM)\cite{OBrien08PNAS, Liu12JPCB}. In the presence of a denaturant of concentration $[C]$, the effective coarse-grained force field for the protein using MTM is given by
\begin{equation}\label{mtmpot}
E_{CG}(\{ {\bf r} \},[C])  = E_{CG}(\{{\bf r} \},0) + \Delta G_{tr}(\{ {\bf r} \},[C]),             
\end{equation} 
   where $E_{CG}(\{{\bf r} \},0)$ is given by Eq.~\ref{pot}, $\Delta G_{tr}(\{ {\bf r} \},[C])$ is the protein-denaturant interaction energy in a solution with denaturant concentration [C], and is given by
\begin{equation}
\Delta G_{tr}(\{ {\bf r} \},[C])  =  \sum_{k=1}^{N} \delta g_{tr,k}([C]) \alpha_k(\{ {\bf r} \})/\alpha_{Gly-k-Gly},
\label{trans_ene}
\end{equation}
where $N$(=$N_{res} \times 2 = 128$) is the number of beads in coarse-grained Protein L, $\delta g_{tr,k}([C])$ is the  transfer free energy of bead $k$, $\alpha_k(\{ {\bf r} \})$ is the solvent accessible surface area (SASA) of the bead $k$ in a protein conformation described by positions  $\{\bf r\}$, $\alpha_{Gly-k-Gly}$ is the SASA of the bead $k$ in the tripeptide $Gly-k-Gly$. The radii for amino acid side chains to compute $\alpha_k(\{ {\bf r} \})$ are given in Table~S2 in Ref.\cite{Reddy15JPCB}. The experimental\cite{Auton04Biochem,OBrien08PNAS,OBrien09Biochem} transfer free energies $\delta g_{tr,i}([C])$, which depend on the chemical nature of the denaturant, for backbone and side chains are listed in Table S3 in Ref.\cite{Liu11PNAS}.  The values for $\alpha_{Gly-k-Gly}$ are listed in Table S4 in Ref.\cite{Liu11PNAS}. 

\bigskip

{\bf Simulations:} The SOP-SC model of the polypeptide chain is simulated using Langevin dynamics at different temperatures ranging from 300 K to 430 K in low friction using the energy function given by eq.~\ref{pot} to compute the average thermodynamic properties of the protein. The equations of motion are integrated using the equation
\begin{equation}
 m \ddot{\vec{r_i}} = -\zeta \dot{\vec{r_i}} + \vec{F_c} + \vec{\Gamma},
\label{Eq. Langevin_dynamics}
\end{equation}
where $m$ is the mass of a protein beads, $\zeta$ is the friction coefficient, $\vec{r_i}$ is the position of the bead $i$, $\vec{F_c} = -\frac{\partial E_{TOT}}{\partial {\vec{r_i}}}$, $\vec{\Gamma}$ is the random force with a white noise spectrum. The autocorrelation function of the random force in the discretised form is given by $\left < \Gamma (t) \ \Gamma (t+nh) \right > = \frac{2 \zeta k_B T}{h} \delta_{0,n}$, where $n = 0, 1, ...$ and $\delta_{0,n}$ is the Kronecker delta function. The Langevin equation is integrated using the velocity Verlet algorithm\cite{Veitshans97FoldDes, Swope82JCP}. We used $\zeta = 0.05 \ m/\tau_L$ and $h = 0.005 \ \tau_L$, where $\tau_L$ is the unit of time  used to advance the simulation. 

To compute thermodynamic properties of the protein in a denaturant solution of concentration $[C]$, $\Delta G_{tr}(\{ {\bf r} \},[C])$  is treated as perturbation to $E_{CG}(\{{\bf r} \},0)$ in Eq.~\ref{mtmpot}, and  Weighted Histogram Method\cite{Kumar92JCompChem,OBrien08PNAS, Liu12JPCB} is used to compute average value of various physical quantities at any [C]. The average value of a physical property $A$, at temperature $T$, and denaturant concentration [C] is computed using the equation

\begin{equation}
\langle A([C],T) \rangle = Z([C],T)^{-1} \sum\limits_{k=1}^{R}\sum\limits_{t=1}^{n_k}\frac{A_{k,t} e^{-(E_{k,t}(\{ {\bf r}_{k,t}\},[0])+\Delta G_{tr}(\{ {\bf r}_{k,t}\},[C]))/k_B T}}{\sum\limits_{m=1}^{R} n_m e^{f_m - E_{k,t}(\{ {\bf r}_{k,t}\},[0])/k_B T_m}},
\label{mtmeq}
\end{equation}
where $R$ is the number of simulation trajectories, $n_k$ is the number of protein conformations from the $k^{th}$ simulation, $A_{k,t}$ is the value of the property of the $t^{th}$ conformation from the $k^{th}$ simulation, $T_m$ and $f_m$ are the temperature and free energy respectively from the $m^{th}$ simulation, $E_{k,t}(\{ {\bf r}_{k,t}\},[0])$ and $\Delta G_{tr}(\{ {\bf r}_{k,t}\},[C]))$ are the internal energy at $[C] = 0$ and MTM energy respectively of the $t^{th}$ conformation from the $k^{th}$ simulation, and $Z([C],T)$ is the partition function given by

\begin{equation}
Z([C],T) = \sum\limits_{k=1}^{R}\sum\limits_{t=1}^{n_k}\frac{ e^{-(E_{k,t}(\{ {\bf r}_{k,t}\},[0])+\Delta G_{tr}(\{ {\bf r}_{k,t}\},[C]))/k_B T}}{\sum\limits_{m=1}^{R} n_m e^{f_m - E_{k,t}(\{ {\bf r}_{k,t}\},[0])/k_B T_m}}.
\label{mtmeqpart}
\end{equation}

We performed Brownian dynamics simulations with the full Hamiltonian given by Eq.~\ref{mtmpot}, and a friction coefficient, which approximately corresponds to that of water to study the burst-phase folding kinetics of Protein L. The equations of motion are integrated using the Ermak-McCammon algorithm\cite{Ermak78JChemPhys}, $\vec{r_i}(t+h) = \vec{r_i}(t) + \frac{h}{\zeta} \vec{F_c} + \vec{\Gamma}$. Here $\vec{\Gamma}$ is a random displacement with a Gaussian distribution with mean zero and variance $\left < \Gamma(h)^2 \right > = \frac{2k_B T h}{\zeta}$. The friction coefficient $\zeta = 31.2 \ m/\tau_H$ approximately corresponds to the value in water and,  the value of $h$ varies from $0.001 \ \tau_H$ to $0.01 \ \tau_H$ depending on the denaturant concentration. In the simulations, the characteristic unit of length $a= 1$ \AA, energy $\epsilon= 1 \ kcal/mole$, and mass $m=1.8\times10^{-22} \ g$ (typical mass of the bead). The unit of time in Langevin dynamics simulations is $\tau_L(= \sqrt{m a^2/\epsilon}) =1.3 \ ps$. In Brownian dynamics, simulation time is mapped into real time, $\tau_H$ using  $\tau_H \approx \frac{\zeta_H a^2}{k_B T} = \frac{(\zeta_H \tau_L/m) \epsilon}{k_B T} \tau_L \approx 47 \ ps$.

\bigskip

{\bf Transition State Analysis:}
108 putative transition state structures (TSS) from the Langevin dynamics trajectory at $T_M=374.5 \ K$ are identified using the conditions 14 \AA $\leq R_g \leq$ 16.2 \AA \ and 0.5 $\leq \chi \leq$ 0.6 (Fig.~S2) for the $P_{fold}$ analysis. To compute $P_{fold}$ for each putative TSS, 500 short simulation trajectories each of $0.15 \ \mu s$ in length are initiated using the putative TSS as the initial conformation to compute the fraction of the trajectories, which land up in the NBA or the UBA (Fig.~\ref{trans}). 

\newpage

\bibliography{Proteins12}
\bibliographystyle{unsrt}

\newpage

\begin{figure} 
\includegraphics[width=6.0in]{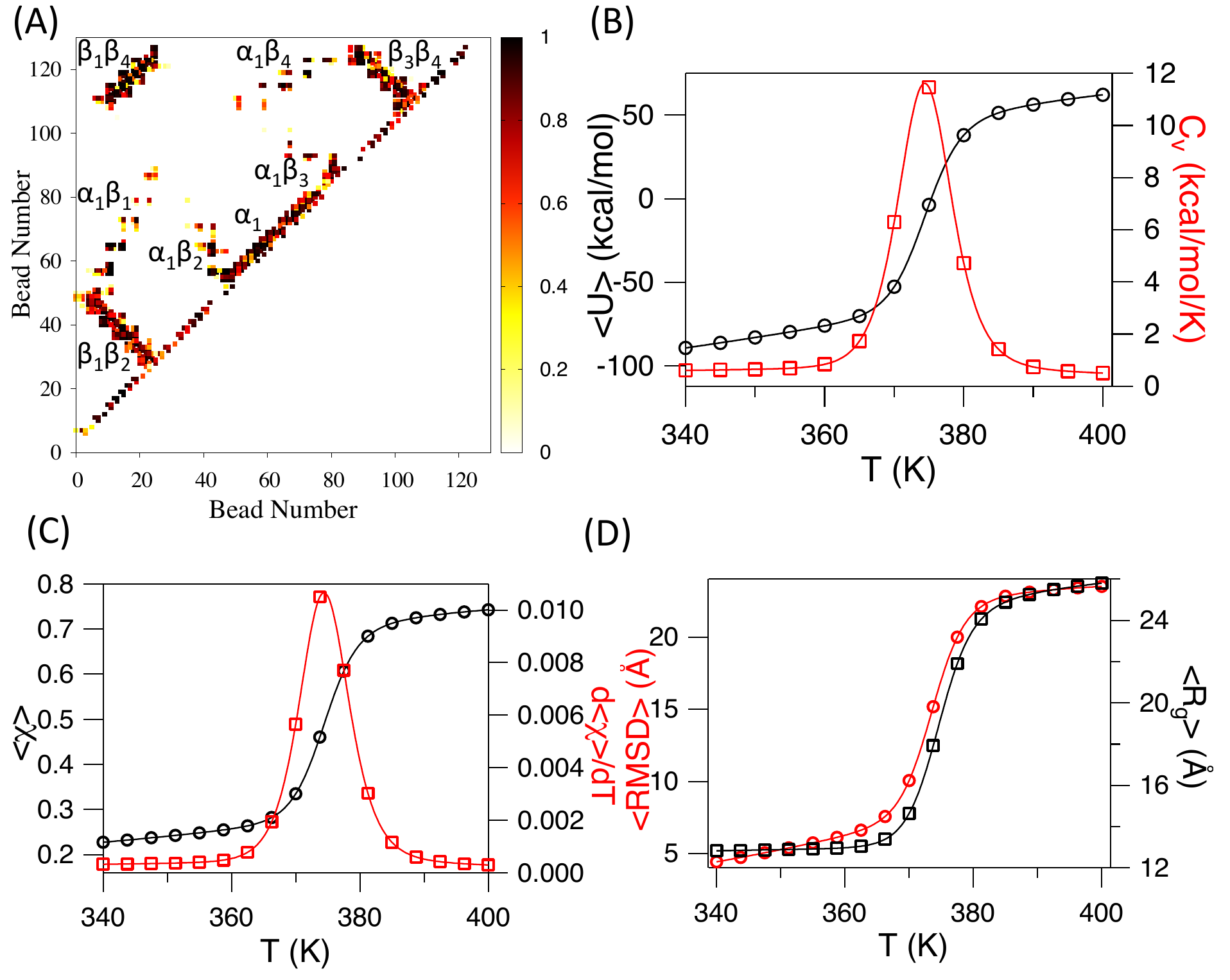}
\caption{(A) The contact map of Protein L shows contacts between various secondary structural elements in the folded state. (B) Average internal energy, $\langle U \rangle$ (empty circles in black), and heat capacity, $C_v$ (empty squares in red), as a function of temperature, $T$. (C) Structural overlap factor, $\langle \chi \rangle$ (empty circles in black), and $d\langle\chi\rangle/dT$ (empty squares in red), as a function of $T$. (D) Root mean square deviation, $\langle RMSD \rangle$ (empty circles in red), and Radius of gyration, $\langle R_g \rangle$ (empty squares in black), as a function of $T$.}\label{temp}
\end{figure}

\begin{figure} 
\includegraphics[width=3.3in]{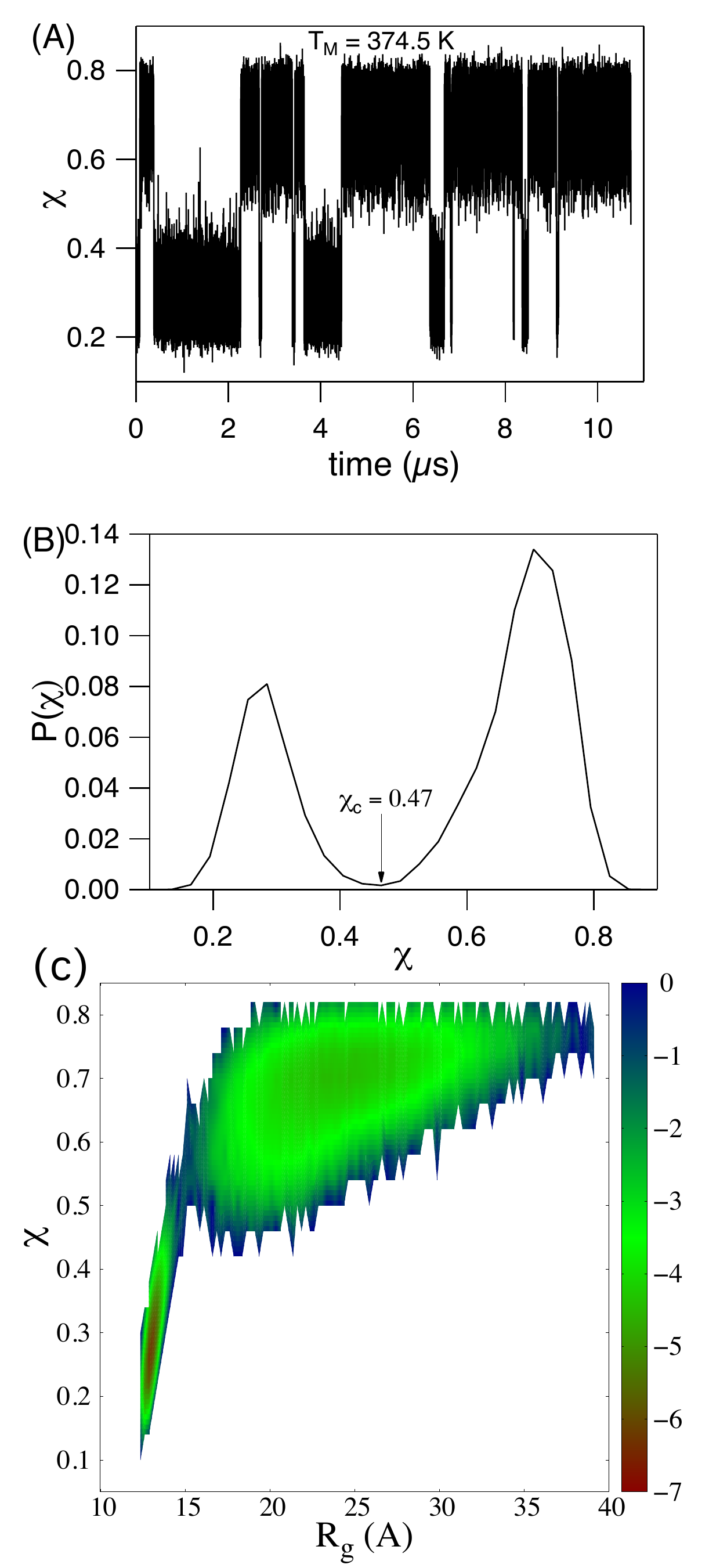}
\caption{(A) Structural overlap factor, $\chi$, plotted as a function of time at the melting temperature, $T_M=374.5 \ K$. (B) Probability distribution of $\chi$, $P(\chi)$, at $T_M$. The value $\chi_c=0.47$ separates the unfolded basin of attraction (UBA) and native basin of attraction (NBA). (C) The free energy projected onto $\chi$ and $R_g$ using the relation, $\Delta G = -k_B T_M \ln(P(R_g,\chi))$, where $P(R_g,\chi)$ is joint probability distribution of $R_g$ and $\chi$ at $T_M$, and $k_B$ is the Boltzmann constant. The two basins corresponding to the UBA and NBA show two-state behaviour at $T_M$.}\label{trap}
\end{figure}

\begin{figure} 
\includegraphics[width=5in]{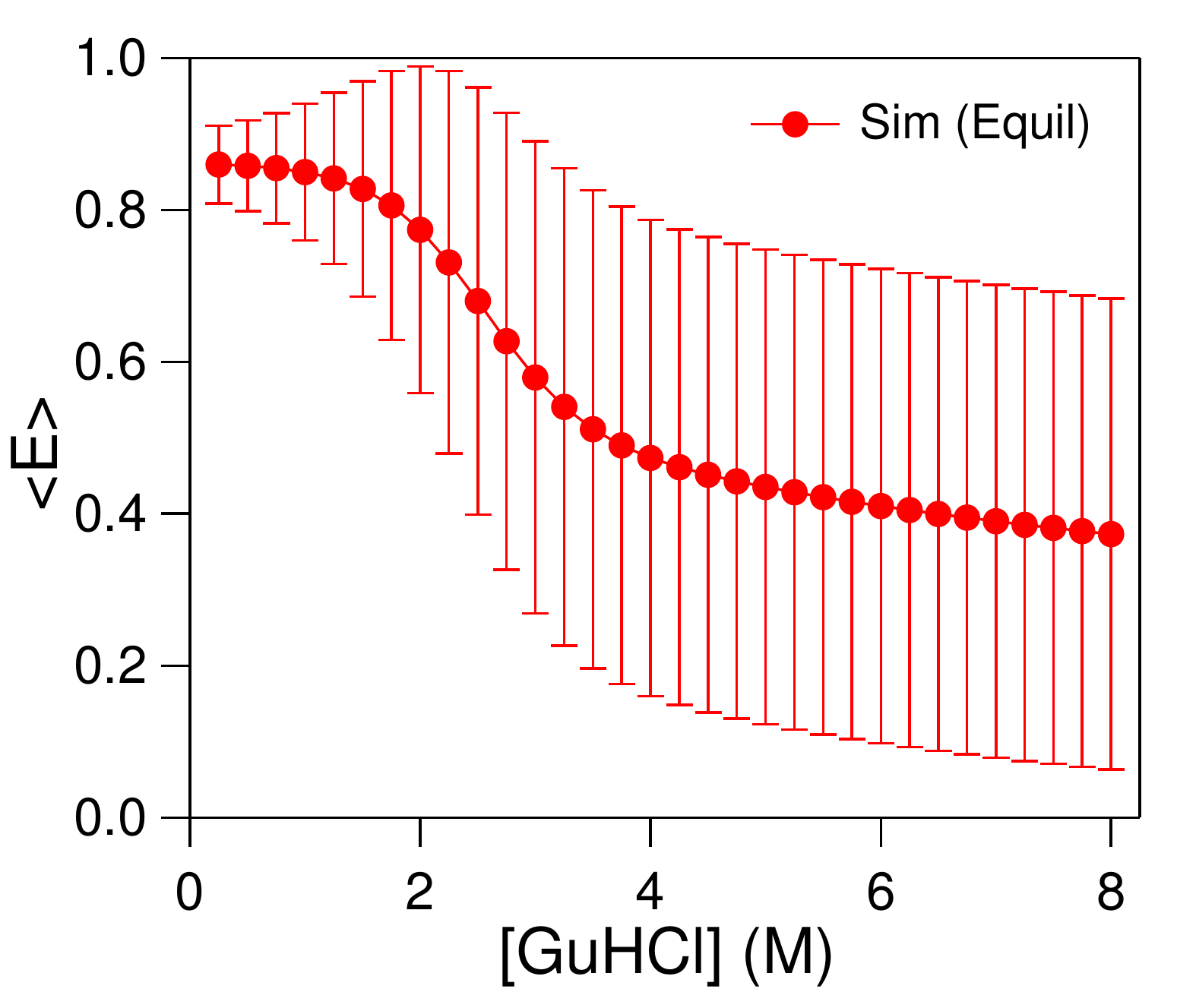}
\caption{Average FRET efficiency, $\langle E \rangle$, as a function of $[GuHCl]$ at $T=357.7 \ K$.}\label{fret_err}
\newpage
\end{figure}

\begin{figure} 
\includegraphics[width=5in]{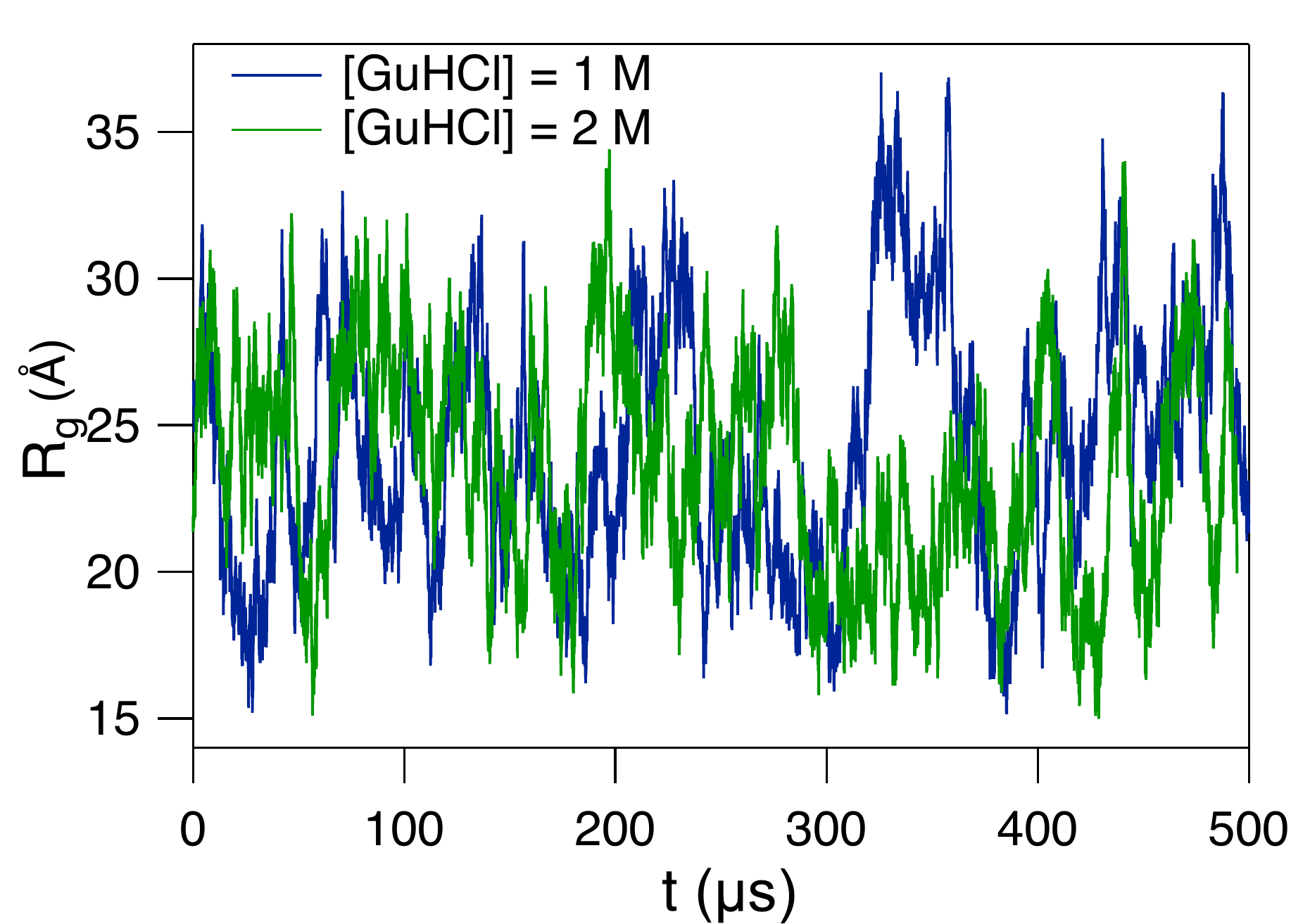}
\caption{Radius of gyration, $R_g$, plotted as a function of time, $t$, for Protein L folding trajectories in $[GuHCl]=1 \ M$ (blue) and $2 \ M$ (green) at $T=357.7 \ K$.}\label{traj_rg}
\newpage
\end{figure}

\begin{figure} 
\includegraphics[width=4in]{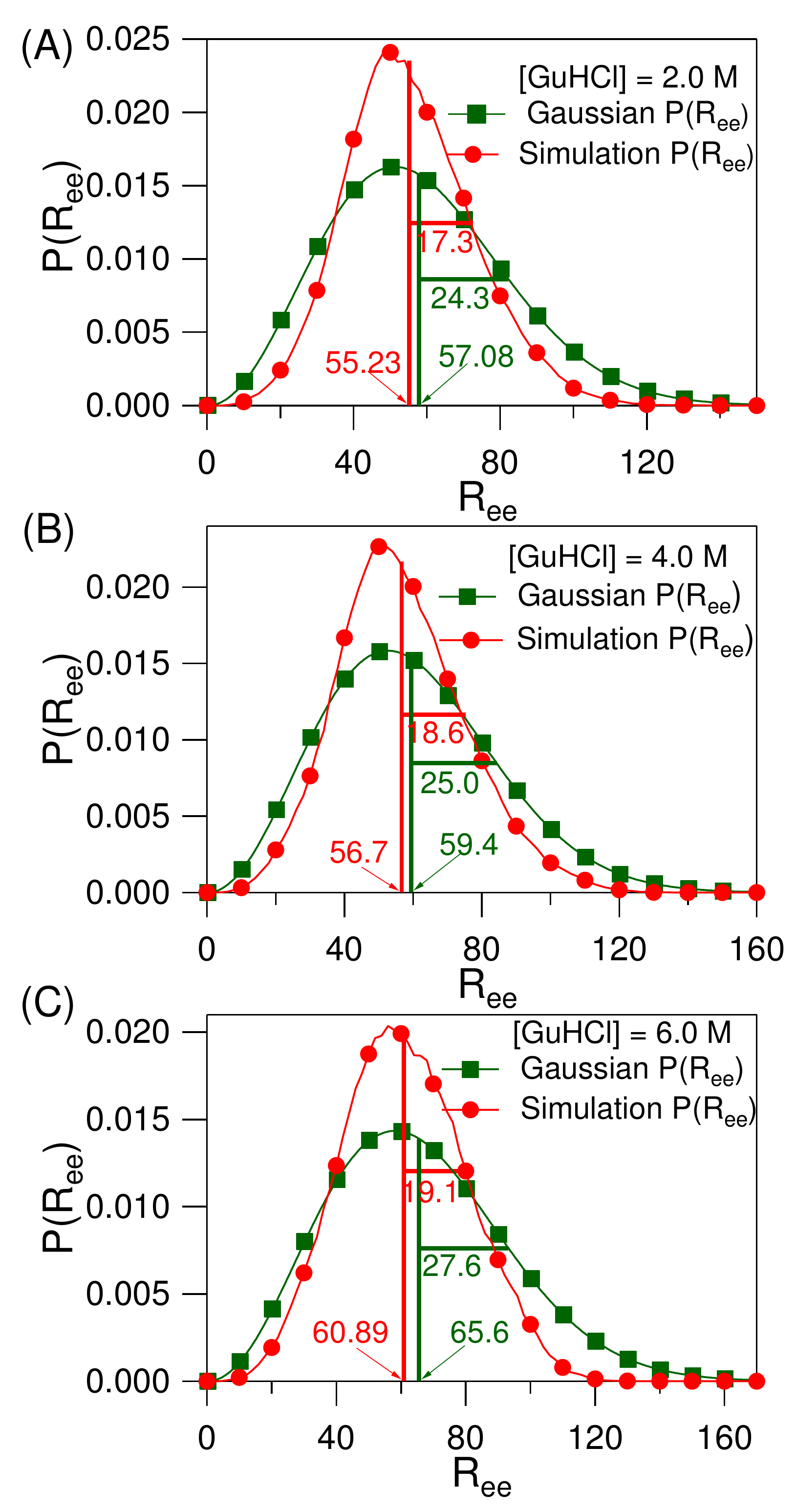}
\caption{The end-to-end distance, $R_{ee}$, probability distribution function $P(R_{ee})$ during the burst phase (initial $0.25 \ ms$) of protein L folding is in red circles. $P(R_{ee})$ estimated from $\langle E^{Burst} \rangle$ and Guassian polymer chain statistics in green squares. (A) $[GuHCl]=2.0 \ M$, $T=357.7 \ K$  (B) $[GuHCl]=4.0 \ M$, $T=357.7 \ K$ and (C) $[GuHCl]=6.0 M$, $T=357.7 \ K$}\label{PRee} 
\newpage
\end{figure}

\begin{figure} 
\includegraphics[width=5in]{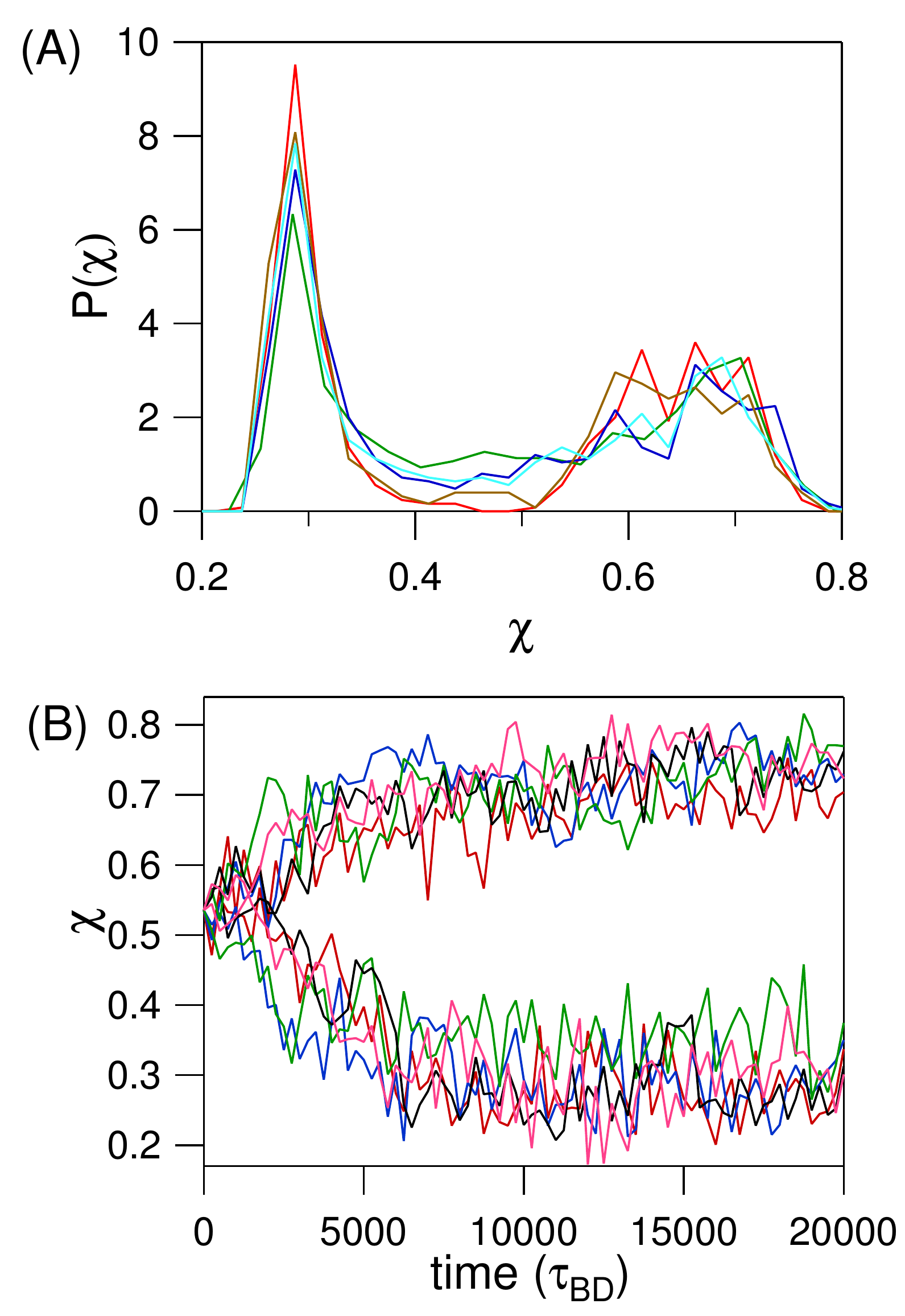}
\caption{(A) The distribution of the final structural overlap factor, $\chi$, for each transition state structure at the end of $0.15 \ \mu s$ computed from 500 simulation trajectories. Data for 5 different structures is shown. (B) Simulation trajectories spawned using a transition state structure as the starting conformation land up in UBA and NBA.}\label{trans}
\newpage
\end{figure}

\begin{table}[ht]
\caption{Parameters for the SOP-Side Chain model}
\label{par}
\begin{center}
\begin{tabular}{|c | c|}
\hline
\hline
 Parameters & Protein \\
\hline
\hline
$R_o$ & 2.0 \AA \\
$k$ & 20 kcal/mol/\AA$^2$ \\
$R_c$ & 8 \AA  \\
$\epsilon_h^{bb}$ & 0.45  kcal/mol  \\
$\epsilon_h^{bs}$ & 0.45  kcal/mol  \\
$\epsilon_l$ & 1.0  kcal/mol  \\
$\sigma^{bb}$ & 3.8 \AA  \\
\hline
\hline
\end{tabular}
\end{center}
\end{table}